\documentclass[journal]{IEEEtran}
\usepackage{cite}
\usepackage{amsmath,amssymb,amsfonts}
\usepackage[amsmath,thmmarks]{ntheorem}
\usepackage{algorithmic}
\usepackage{graphicx} 
\usepackage{subcaption}
\usepackage{textcomp}
\usepackage{xcolor}

\newenvironment{proof}{{\indent \indent \it Proof:}}{\hfill $\blacksquare$\par}
\usepackage{lettrine}

{\theoremheaderfont{\bfseries}
	\theorembodyfont{\upshape}
	\theoremseparator{:}
	\newtheorem{proposition}{\hspace{1em}Proposition}}

{\theoremheaderfont{\bfseries}
	\theorembodyfont{\upshape}
	\theoremseparator{:}
	}

{\theoremheaderfont{\bfseries}
	\theorembodyfont{\upshape}
	\theoremseparator{:}
	}

{\theoremheaderfont{\bfseries}
	\theorembodyfont{\upshape}
	\theoremseparator{:}
	}

{\theoremheaderfont{\bfseries}
	\theorembodyfont{\upshape}
	\theoremseparator{:}
	\newtheorem{remark}{\hspace{1em}Remark}}
\usepackage{indentfirst}
\usepackage{array}
\usepackage{bm}
\usepackage{textcomp}
\usepackage{stfloats}
\usepackage{url}
\usepackage{verbatim}
\usepackage{graphicx}
\usepackage{caption}
\usepackage{amssymb}
\usepackage{float}
\usepackage{balance}
\usepackage{diagbox}
\usepackage{booktabs}

\allowdisplaybreaks[4]

\makeatletter

\newcommand{\Rmnum}[1]{\expandafter\@slowromancap\romannumeral #1@}
\makeatother
\usepackage{hyperref}

\begin{document}

\title{Curved Waveguide-Enabled Pinching-Antenna System (C-PAS): Communication Performance Analysis}

\author{
Yayun Qu, Kunrui Cao, \textit{Senior Member, IEEE}, Tao Wang, Lu Lv, \textit{Member, IEEE}, Jiwei Tian, \\  Dimitrios Tyrovolas, \textit{Member, IEEE}, Panagiotis D. Diamantoulakis, \textit{Senior Member, IEEE}, \\ and George K. Karagiannidis, \textit{Fellow, IEEE}

\thanks{Yayun Qu, Kunrui Cao, and Tao Wang are with the Information Support Force Engineering University, Wuhan 430035, China, and also with the School of Information and Communications, National University of Defense Technology, Wuhan 430035, China
(e-mail: quyayun@nudt.edu.cn; krcao@nudt.edu.cn; twang@nudt.edu.cn).\\ 
\indent Lu Lv is with the State Key Laboratory of Integrated Services Networks, Xidian University, Xi'an 710071, China
(e-mail: lulv@xidian.edu.cn).\\ 
\indent Jiwei Tian is  with the School of Cyber Science and Engineering, Xi’an Jiaotong University, Xi’an 710049, China (e-mail: tianjiwei2016@163.com).\\ 
\indent D. Tyrovolas, P. D. Diamantoulakis, and G. K. Karagiannidis are with the Department of Electrical and Computer Engineering,
Aristotle University of Thessaloniki, 54124 Thessaloniki, Greece (e-mail:
tyrovolas@auth.gr; padiaman@auth.gr; geokarag@auth.gr).\\ 

   }
     
}

\maketitle

\begin{abstract}
Existing studies on the pinching-antenna system (PAS)
assume that waveguides are deployed straight, which fails to serve communication regions with curved boundaries. To address this limitation, this paper proposes a curved waveguide-enabled pinching-antenna system (C-PAS), where the waveguide is placed along the building ceiling in an arc to maximize the line-of-sight (LoS) coverage. On this basis, the optimal pinching‑antenna (PA) placement strategy and the nearest PA placement strategy are presented. The optimal strategy distinguishes between scenarios, i.e., with or without inner-wall blockage, and derives a closed-form solution for the optimal PA position to maximize the signal-to-noise ratio (SNR) received at the user. Meanwhile, the nearest strategy aligns the PA with the angular position of a user in polar coordinates, thereby accommodating the curved geometry of the region. Furthermore, the outage probability (OP) and the average rate (AR) are analyzed for each strategy, and the corresponding analytical expressions are derived, respectively. The results show that the optimal PA placement strategy achieves better OP and AR performance than the nearest strategy, particularly under large waveguide loss coefficient or waveguide height. Moreover, for a service region of fixed area, there exists an optimal sector angle or inner-wall radius that either minimizes the outage probability or maximizes the average rate. Furthermore, the choice of a waveguide bending radius is influenced by the transmit power of the base station, where a smaller bending radius is preferable at high power and the middle arc performs best at low power.
\end{abstract}


\section{Introduction}
\lettrine[lines=2, loversize=0.15]{T}{he} sixth-generation (6G) communication networks pose stringent requirements for higher capacity, lower latency, and enhanced reliability. However, path loss and blockage in signal propagation are extremely prominent in the 6G high-frequency bands, thereby constituting a key barrier to realizing these requirements \cite{Cai2025NextGenTRX, lu2024tutorial,Wang2023Road6G,cui2023near}. To this end, reconfigurable intelligent surface (RIS), fluid antenna (FA), and mobile antenna (MA) have been proposed gradually\cite{chen2025secure,wang2026pentahedral}. While these technologies effectively boost system capacity and coverage by improving the channel environment, they face practical deployment limitations. 
Specifically, RIS employs two separate links, i.e., the transmitter‑to‑RIS link and the RIS‑to‑receiver link, via phase‑shift optimization to effectively enhance coverage and capacity. However, the increased propagation distance results in severe multiplicative fading \cite{chen2025double,cao2026self,cao2026reliable}. FA/MA can attain additional spatial degrees of freedom through flexible position adjustments, thereby helping to improve channel capacity. Nevertheless, the movement of antenna elements is confined to a wavelength scale, which generally cannot overcome the effects of large‑scale fading. In addition, once these systems are deployed, it becomes difficult to change the number of antenna elements, reflecting insufficient deployment flexibility \cite{Hong2026FASsurvey,New2024FASoutage, pang2026secure, Zhu2025Movable}.

In recent years, the pinching-antenna system (PAS) has attracted widespread attention and has been extensively studied as an emerging flexible antenna architecture \cite{Fukuda2022Pinching}. PAS employs a low-loss dielectric waveguide to transmit high-frequency signals and activates radiation points at arbitrary positions along the waveguide via a pinching mechanism. PAS enables the antenna to be dynamically deployed in proximity to users, thereby significantly shortening the free-space propagation distance and allowing rapid relocation when encountering blockages so as to maintain a line-of-sight (LoS) communication link \cite{Liu2026pin,yang2025pinching,wang2025modeling,DT2026HOW}.

In order to realize the full potential of PAS, waveguide loss is a critical factor that must be considered in practical deployment. In this context, recent studies have investigated this issue from various perspectives. 
In terms of hardware characterization, the authors in\cite{zhang2025directional} adopted a measured power loss coefficient (1.3 dB/m) in directional PAS to analyze the trade-off between waveguide loss and free-space reliability, and identified the conditions under which an optimal position exists. 
Regarding fundamental modeling and performance analysis, the waveguide loss coefficient was introduced into the derivation of outage probability (OP) and average rate (AR)  in \cite{tyrovolas2026performance}, revealing significant performance degradation over long waveguides.
Similarly, the authors in\cite{xu2026attenuation} provided a closed-form solution for the optimal pinching-antenna (PA) position in a single-user scenario and quantified the rate loss caused by ignoring attenuation. 
Extending the analysis to different deployment scenarios, a circular indoor environment was investigated, and it was found that under partial coverage with lossy waveguides, system performance varies non-monotonically with waveguide length, with the optimal length decreasing as the loss coefficient increases \cite{Zhang2026iot}.
In addition, the authors in\cite{Xu2026LoS} combined waveguide attenuation with probabilistic LoS blockage and derived the ergodic rate loss due to neglecting attenuation, showing that the loss is significant under sparse blockage yet negligible under dense blockage.

Apart from performance analysis, researchers have also focused on mitigating waveguide loss through structural adjustments and algorithmic optimizations.
At the structural level, the authors in\cite{ouyang2025swans} proposed segmented waveguide PAS (SWANs), which divide a long waveguide into multiple short segments with independent feeding, thereby reducing the average distance from the PA to the feeding point and thus lowering the loss.
At the algorithmic level, various optimization strategies have been developed. The authors in\cite{shan2025multicast} optimized the PA position via element-wise search in a multi-group multicast scenario to alleviate the constraint of waveguide loss on the worst-user rate. In \cite{pakravan2025ai}, waveguide loss was combined with a non-linear energy harvesting model, and deep reinforcement learning was employed to dynamically optimize the PA position so as to compensate for energy loss.
Beyond single-waveguide scenarios, the work in \cite{mao2025isac} jointly optimized the PA position and beamforming through the successive convex approximation (SCA) in a multi-waveguide integrated sensing and communication (ISAC) system, and revealed that increasing the number of waveguides reduces the average distance and thus mitigates the loss.

Furthermore, to cope with non‑ideal channel conditions, researchers have developed robust algorithms that explicitly account for waveguide loss.
To counter eavesdropping threats, a 3D blockage‑aware channel model was constructed, based on which beamforming, artificial noise, and PA position were jointly optimized to suppress eavesdropping while preserving a LoS link for legitimate users \cite{zhao2026secure}.
Separately, to handle channel uncertainty, the authors in\cite{sun2026robustmulti} investigated a multi-user PAS, aiming to minimize the transmit power under channel uncertainty while guaranteeing the rate requirements of all users in the worst case. The work in \cite{Sun2026RobustSingle} studied a single-user PAS and maximized the worst-case achievable rate under channel uncertainty, taking into account both lossy and lossless waveguide transmissions. Both works demonstrate that the worst-case performance of PAS under imperfect channel state information (CSI) still outperforms that of fixed-antenna systems under perfect CSI.

\subsection{Motivation and Contributions}
\begin{figure}[t]
    \centering
    \begin{subfigure}[b]{0.49\linewidth}
        \includegraphics[width=\linewidth]{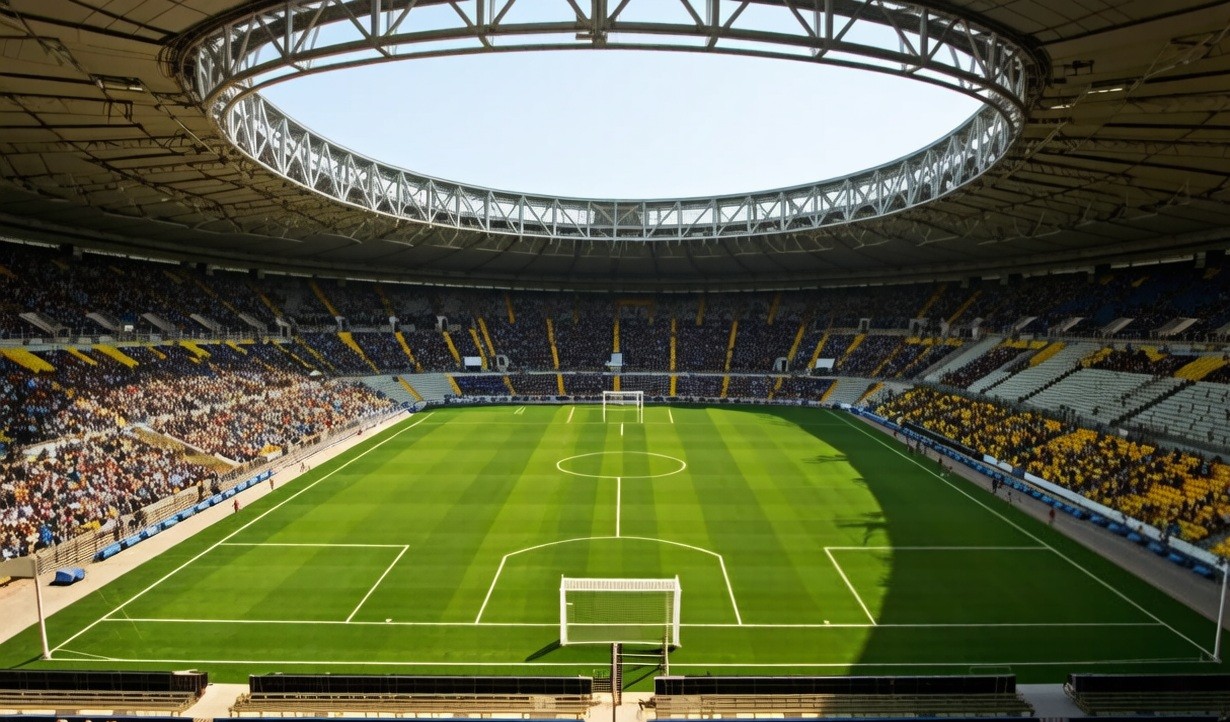}
        \label{fig:sub1}
    \end{subfigure}
    \hfill
    \begin{subfigure}[b]{0.49\linewidth}
        \includegraphics[width=\linewidth]{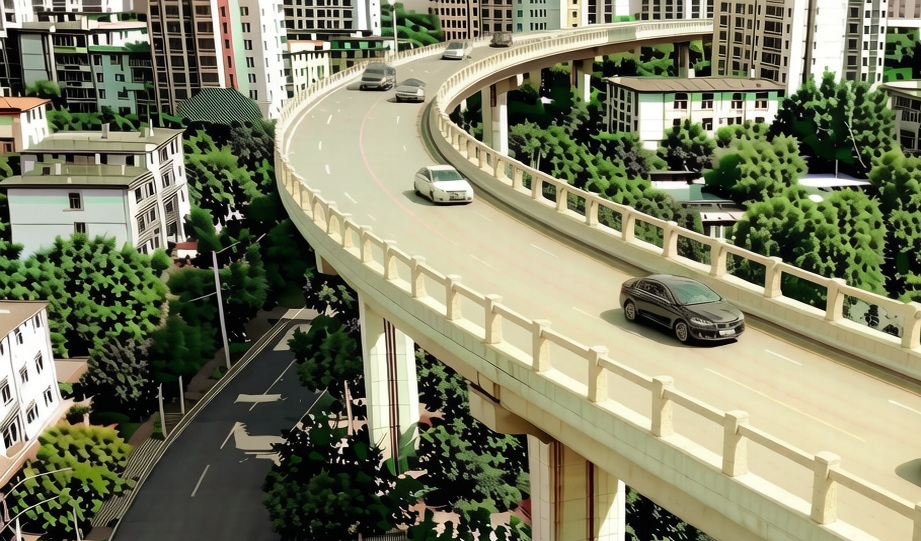}
        \label{fig:sub2}
    \end{subfigure}
    \\
    \begin{subfigure}[b]{0.49\linewidth}
        \includegraphics[width=\linewidth]{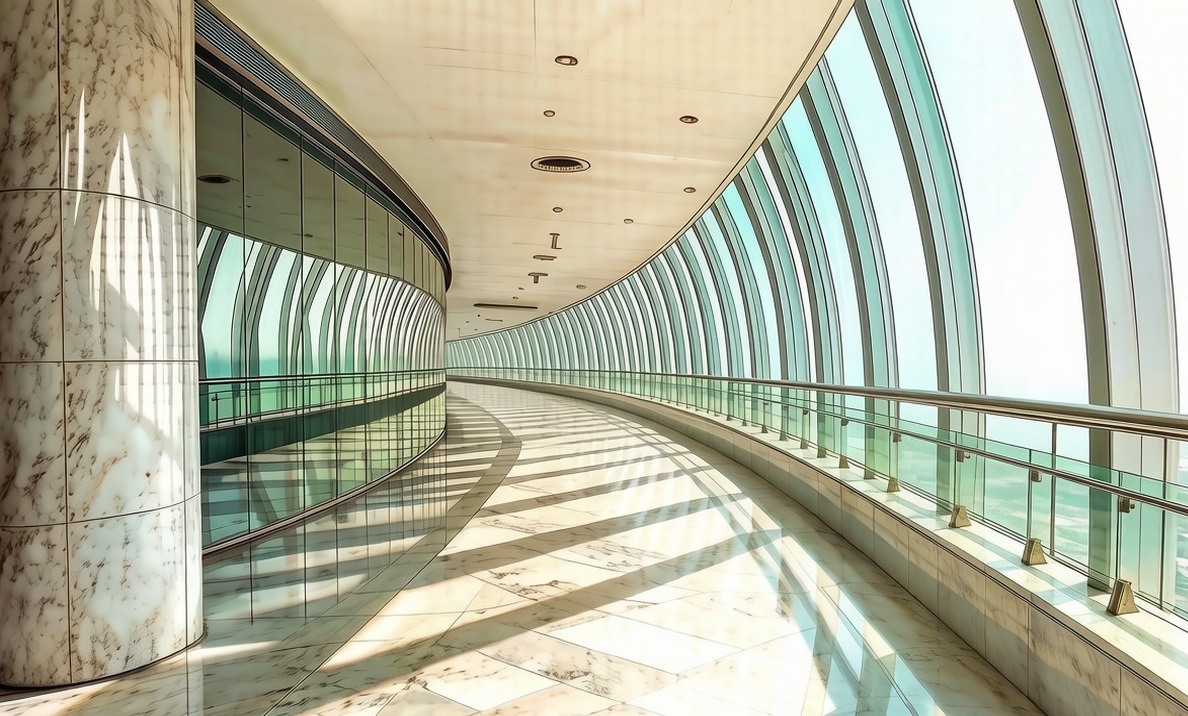}
        \label{fig:sub3}
    \end{subfigure}
    \hfill
    \begin{subfigure}[b]{0.49\linewidth}
        \includegraphics[width=\linewidth]{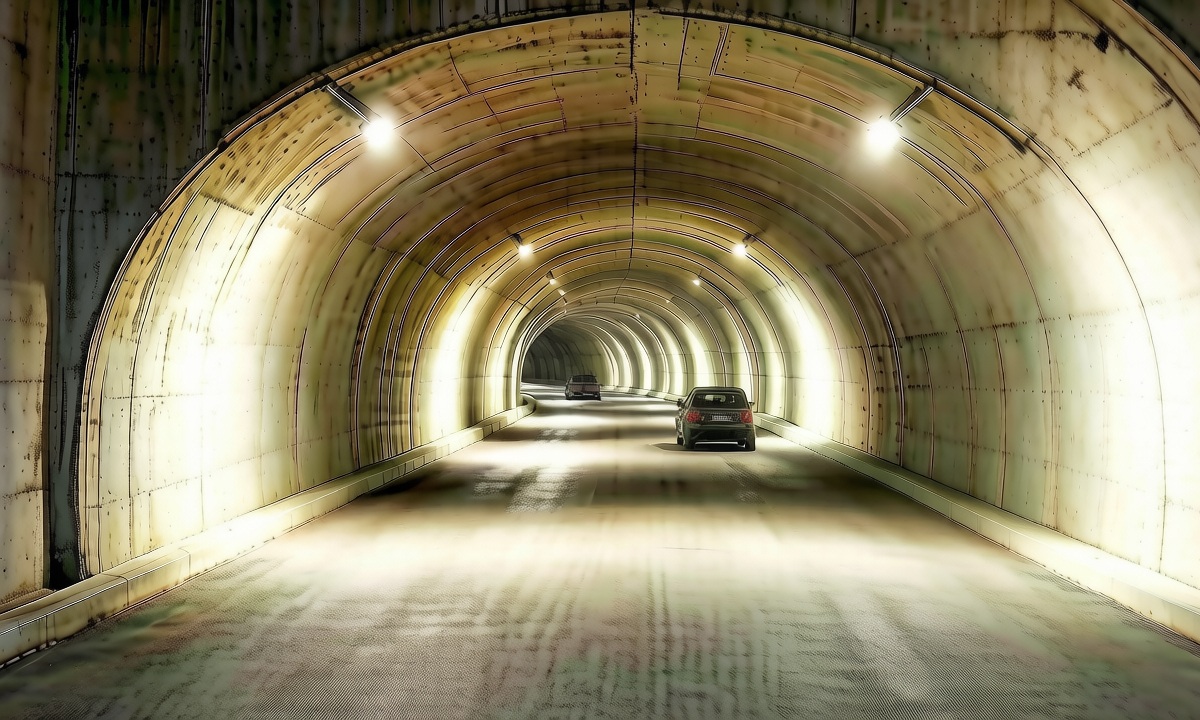}
        \label{fig:sub4}
    \end{subfigure}
    \caption{Typical scenarios where straight waveguide deployment is infeasible}
    \label{fig:fourgrid}
\end{figure}

All the existing studies on PAS consider straight line waveguide deployment, overlooking the inevitable bending of waveguides caused by building structures, terrain, and obstacles in actual environments. As shown in Fig.~\ref{fig:fourgrid}, practical communication scenarios often include non-rectangular service areas, such as domed stadiums, elevated loop ramps, arc-shaped corridors, and curved tunnels. Since the boundary of the service area is curved with a large curvature, traditional straight line waveguides, whether deployed radially or tangentially, cannot achieve full coverage of the antennas, resulting in the failure of the straight waveguide channel model.
The waveguides need to be deployed in a curved manner along the tangential direction to ensure that any location within the service area is covered by the PA radiation. Unlike the random blockages in conventional rectangular scenarios, the arc-shaped service area inherently features an inner-wall blockage, which permanently restricts the LoS communication range for users at certain positions and thus must be taken into account in PA deployment. Moreover, the bending radius of the curved waveguide affects the signal coverage and the deployment of the PA, becoming a new issue that necessitates investigation.

Based on the above analysis, the existing straight-waveguide model can no longer cope with the challenges posed by curved areas. To fill this gap, this paper proposes a curved waveguide-enabled pinching-antenna system (C-PAS). The main contributions of this paper are summarized as follows.
\begin{itemize}
\item We propose the C-PAS communication architecture, which deploys curved waveguides along the building ceiling to serve arc‑shaped service regions.  Based on the geometric characteristics of arc‑shaped regions, an equal‑width annular sector communication model is established. In addition, the LoS blockage condition caused by the inner wall is analyzed, and a mathematical criterion for LoS communication is derived.

\item The optimal and nearest PA placement strategies in C‑PAS are proposed. Specifically, the optimal strategy distinguishes between scenarios with and without inner-wall blockage and derives a closed-form expression for the optimal PA position to maximize the signal‑to‑noise ratio (SNR) received at the user. The nearest strategy serves as a low-complexity benchmark that aligns the PA with the angular position  of a user  in polar coordinates, thus avoiding inner-wall blockage.

\item The performance under various system parameter configurations is evaluated, and analytical expressions for the OP and AR are derived, respectively. All theoretical results are validated by Monte Carlo simulations, providing design guidelines for practical application and deployment.

\item 
Theoretical and numerical results show that 1) The OP and AR of the optimal PA placement strategy are superior to those of the nearest placement strategy, especially for large waveguide loss coefficient or waveguide height.
2) For a service region of fixed area, there exists an optimal sector angle or inner-wall radius that either minimizes the outage probability or maximizes the average rate.
3) At high transmit power of the base station, a smaller waveguide bending radius is preferable, while at low power,  the middle arc achieves the best performance.
4) For small waveguide loss coefficient or low waveguide height, the waveguide bending radius should be chosen as the middle arc.
5) For high waveguide height  or large loss coefficient, a smaller radius of the curved waveguide  is preferable.
\end{itemize}
\subsection{Organization}
The remainder of this paper is organized as follows. Section II presents the C‑PAS model and the SNR expression at the user. Section III presents the optimal PA  and nearest PA  placement strategies, and derives the closed‑form expression for the optimal PA position, taking LoS blockage into account. Section IV provides analytical expressions for the OP and the AR. Section V validates the theoretical analysis through simulations and reveals the impact of key parameters on the system performance. Section VI concludes the paper.
\section{System Model}
\begin{figure}[t]
	\centering
	\includegraphics[width=\linewidth]{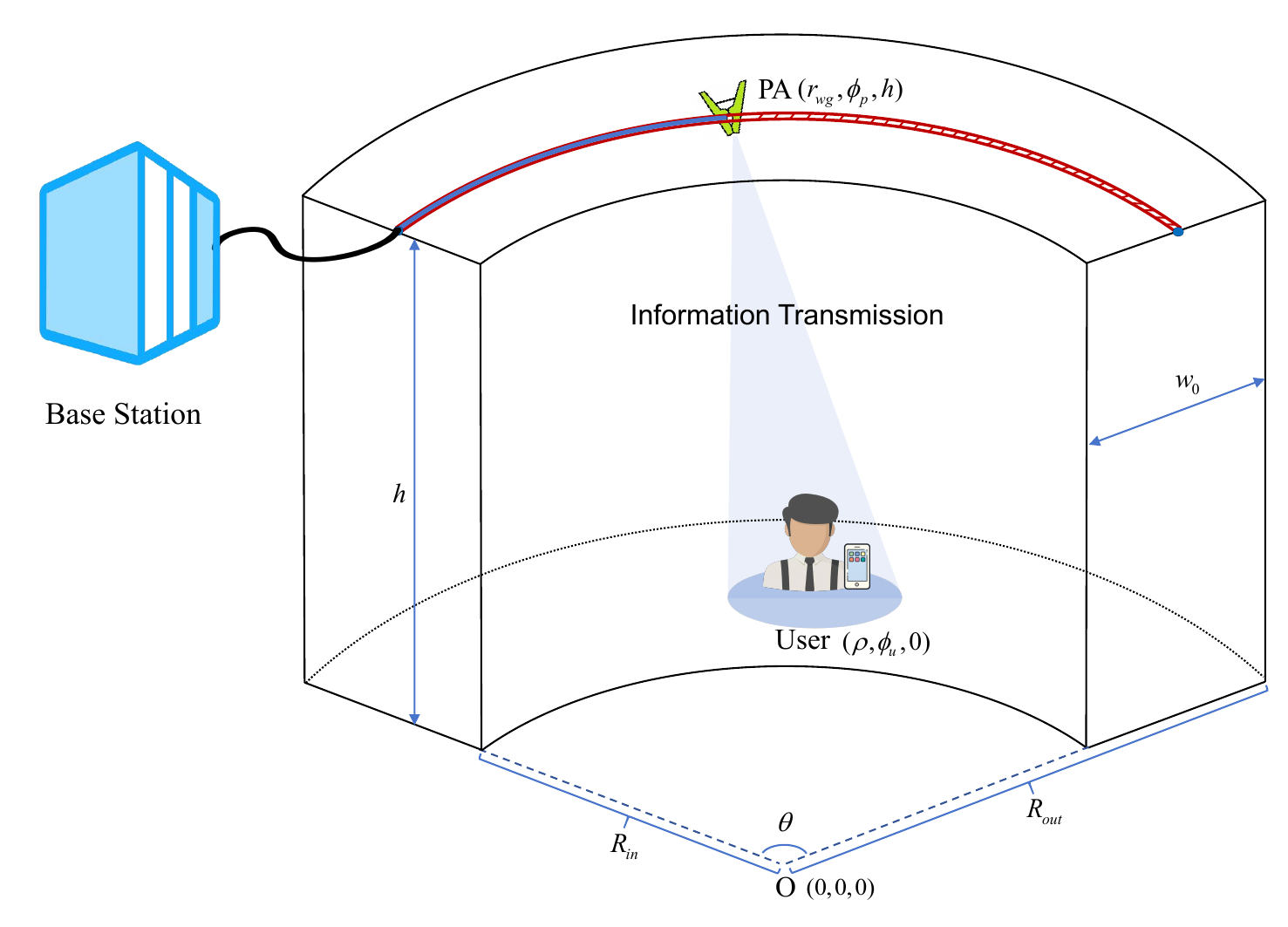}
	\captionsetup{font={footnotesize}} 
	\caption{C-PAS communication model.}
	\label{fig1}
\end{figure}

As shown in Fig.~\ref{fig1}, we consider a downlink wireless communication scenario assisted by the C-PAS, where the communication takes place indoors with curved walls. The C-PAS is deployed on the ceiling and consists of a base station, a dielectric waveguide, and a PA mounted on the waveguide. The signal of the base station is transmitted through the curved waveguide to the PA, which radiates the signal to a user. The user is distributed randomly and uniformly within an arc-shaped communication region of width \(w_0=R_{\text{out}}-R_{\text{in}}\). The position of the user is described in polar coordinates as \((\rho, \phi_u, 0)\), where the radial distance \(\rho \in [R_{\text{in}}, R_{\text{out}}]\) and the angle \(\phi_u \in [0, \theta]\). The middle-arc radius is defined as \(R_{\text{mid}} = (R_{\text{in}} + R_{\text{out}})/2\) to facilitate subsequent analysis.

In arc‑shaped regions, the waveguides are deployed tangentially along the ceiling, as discussed in the introduction. The feed point is located at \((r_{wg}, 0, h)\), where 
\(r_{wg} \in [R_{\text{in}}, R_{\text{out}}]\) is the waveguide bending radius and \(h\) is the ceiling height. 
The PA can be activated at any position on the waveguide, denoted as \((r_{wg}, \phi_p, h)\).

The channel $h_{\mathrm{pu}}$ between the PA and the user in C-PAS can be expressed as \cite{tyrovolas2026performance} 
\begin{align}
	\label{h1_def}
	{h_{\mathrm{pu}}} = \frac{{\sqrt \eta  \, {e^{ - j\frac{{2\pi }}{\lambda }\left\| {{\psi _u} - {\psi _p}} \right\|}}}}{{\left\| {{\psi _u} - {\psi _p}} \right\|}},
\end{align}
where $\eta = \frac{\lambda^2}{16\pi^2}$ represents the path loss at a reference distance of 1 meter, \(\lambda\) is the free-space wavelength, and ${{\left\| {{\psi _u} - {\psi _p}} \right\|}} $ denotes the distance between the PA and the user.
Then the additional phase shift \(h_{\mathrm{ap}}\) from the feed point to the PA can be expressed as \cite{tyrovolas2026performance} 
\begin{align}
	h_{\mathrm{ap}} = e^{-j\frac{2\pi}{\lambda_g}\ell},
\end{align}
where  $\lambda_g = \frac{\lambda}{n_{\text{eff}}}$,  \(n_{\text{eff}}\) is the effective refractive index of the curved waveguide, and $\ell$ is the length of the signal path through the curved waveguide.

The curved waveguide has both an absorption coefficient \(\alpha_{\text{abs}} \in [0, +\infty)\) and a radiation coefficient \(\alpha_{\text{rad}} \in [0, +\infty)\),  characterizing the power loss experienced by the signal during propagation\cite{deck1998,kozyreff2016dispersion}. The radiation coefficient \(\alpha_{\text{rad}} \) is given by\cite{kozyreff2016dispersion}
\begin{equation}
\alpha_{\text{rad}} = \frac{4e^{2\nu S(x)}}{\xi \, r_{wg} \sqrt{n_{wg}^2 - 1}},
\label{eq:bend_loss}
\end{equation}
where $S(x) = \sqrt{1 - x^2} - \ln \left( \frac{1 + \sqrt{1 - x^2}}{x} \right)$ and $x = \frac{k_r r_{wg}}{\nu}$, with $\nu$ the azimuthal number, $n_{wg}$ the refractive index of the waveguide core, $\xi$ the polarization factor, and $k_r$ the real wavenumber.

The received signal at the user is thus given by
\begin{align}
	\label{yr_def}
	y_r = \sqrt{P_t \cdot e^{-\left(\alpha_{\mathrm{abs}} + \alpha_{\mathrm{rad}}\right)\ell}} \cdot h_{\mathrm{pu}} h_{\mathrm{ap}} s + w_n,
\end{align}
where \(s\) is the transmitted signal satisfying \(\mathbb{E}[s^2] = 1\), and \(\mathbb{E}[\cdot]\) denotes the statistical expectation. \(P_t\) is the transmit power of the BS, and \(w_n\) represents additive white Gaussian noise (AWGN) with zero mean and variance of \(\sigma^2\). Thus the signal-to-noise ratio (SNR) received at the user can be written as
\begin{align}
\gamma_r &= \frac{\eta P_t \mathrm{e}^{-\left(\alpha_{\mathrm{abs}} + \alpha_{\mathrm{rad}}\right)\ell} \cdot \left| \mathrm{e}^{-j\left( \frac{2\pi}{\lambda} \|\psi_u - \psi_p\| + \frac{2\pi}{\lambda_g} \ell \right)} \right|^2}{\sigma^2 \|\psi_u - \psi_p\|^2} \nonumber \\
&= \frac{\eta P_t \mathrm{e}^{-\left(\alpha_{\mathrm{abs}} + \alpha_{\mathrm{rad}}\right)\ell}}{\sigma^2 \left\|\psi_u - \psi_p\right\|^2}.\label{gamma_r_combined}
\end{align}
In the curved waveguide of C-PAS, $\ell$ is given by
\begin{align}
\label{eq:L}
\ell = r_{wg} \phi_p .
\end{align}
Accordingly, \eqref{gamma_r_combined} can be rewritten as
\begin{align}
	\label{gamma_r_final}
	\gamma_r = \frac{\eta P_t e^{-\alpha r_{wg}  \phi_p }}{\sigma^2 \left\|\psi_u - \psi_p\right\|^2},
\end{align}
 where $	\alpha= \alpha_{\mathrm{abs}} +\alpha_{\mathrm{rad}}$.

\section{PA placement strategies in C-PAS}
This section proposes two strategies for the C-PAS, i.e., the optimal PA placement strategy and the nearest PA placement strategy. For each strategy, we analyze the SNR received at the user and derive a closed‑form expression for the PA position.

\subsection{Optimal PA Placement Strategy}

To maximize the received SNR of the user, we propose the \textbf{optimal PA placement strategy}, where the PA position that maximizes the received SNR is defined as the optimal PA position. 
 The Euclidean distance between the user and the PA is denoted as \(d\). According to the law of cosines, we have
\begin{align}
	\label{d_def}
	d = \sqrt{\rho^2 + r_{wg}^2 + h^2 - 2\rho r_{wg} \cos(\phi_u - \phi_p)}.
\end{align}
Thus the received SNR of the user in \eqref{gamma_r_final} can be rewritten as
\begin{align}
	\label{gamma_r_arc}
	\gamma_r^{\text{opt}} = \frac{\eta P_t e^{-\alpha r_{wg} \phi_p} / \sigma^2}{\rho^2 + r_{wg}^2 + h^2 - 2\rho r_{wg} \cos(\phi_u - \phi_p)}.
\end{align}

From \eqref{gamma_r_arc}, given a fixed user location and waveguide deployment, the received SNR is maximized by tuning the angular position of PA on the waveguide. In C-PAS, however, the optimal PA placement is examined in two cases, distinguished by whether the LoS link between the PA and the user is blocked by the inner wall.
\subsubsection{Optimal PA Placement without Blockage}
\begin{figure}[t]
	\centering
	\includegraphics[width=\linewidth]{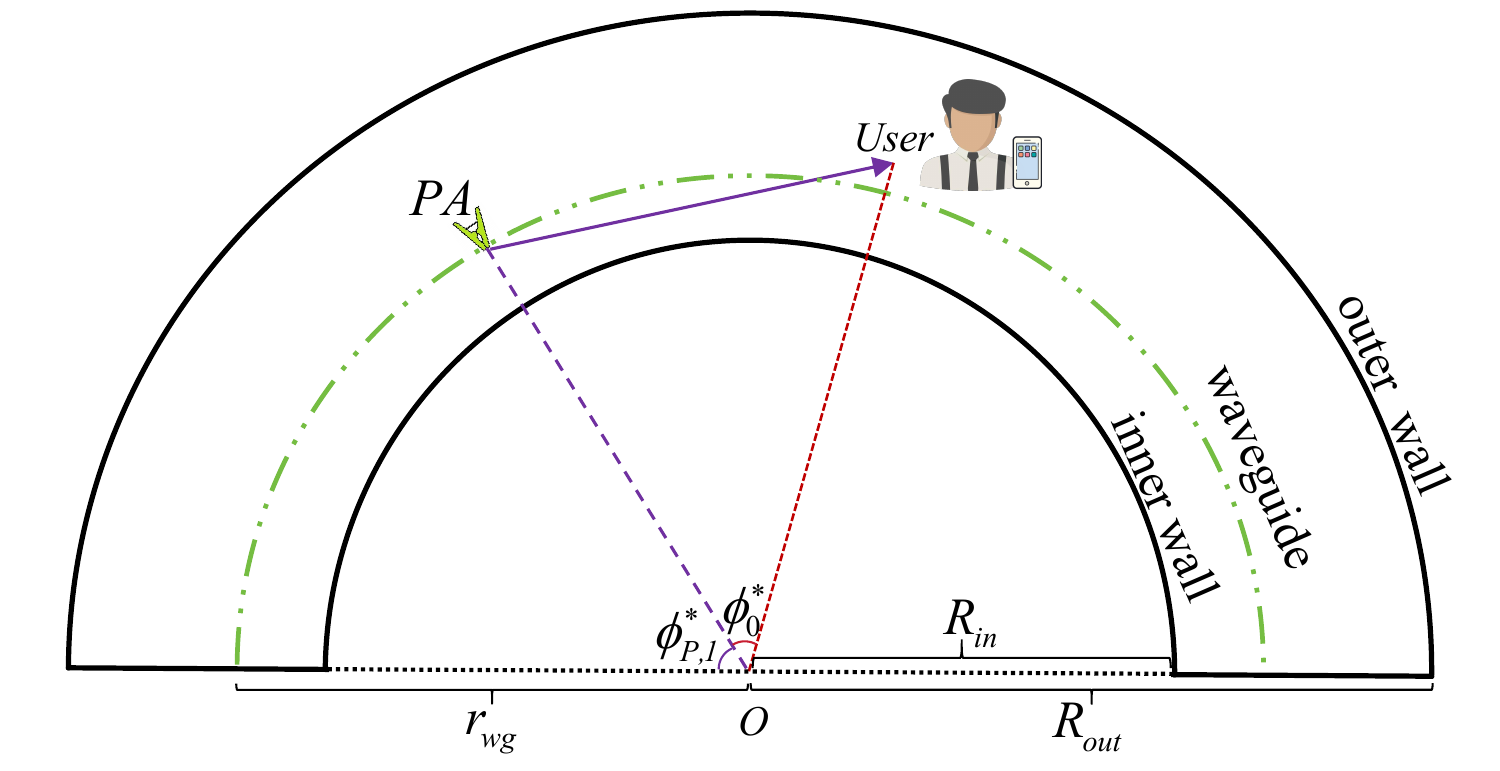}
	\captionsetup{font={footnotesize}} 
	\caption{Planar layout of C-PAS without blockage.}
	\label{fig:no_blockage}
\end{figure}

In the arc-shaped region, the blockage of the LoS communication between the PA and the user arises from the inner wall. Since the height of the curved waveguide does not affect the LoS communication between the PA and the user, we vertically project the system onto the ground for ease of analysis.
Fig.~\ref{fig:no_blockage} illustrates the no-blockage LoS scenario, where the line between the PA and the user avoids the inner wall, thus enabling direct communication.
With waveguide losses accounted for, moving the PA from \( \phi_p = \phi_u \) results in the following two distinct phenomena.
\begin{itemize}
\item When the PA moves toward the waveguide end, $\phi_p$ increases. 
Both the free-space propagation distance and the in-waveguide propagation distance increase, leading to stronger free-space attenuation and higher waveguide loss. Consequently, the overall system performance is degraded.
\item When the PA moves toward the waveguide start, $\phi_p$ decreases. 
The free-space propagation distance becomes longer, whereas the in-waveguide propagation distance becomes shorter. 
As the guided-section loss is reduced, the overall system performance can be improved.
\end{itemize}

Thus the optimal PA angle position without blockage \(\phi_{p,1}^*\) lies between the waveguide start and the user, i.e., \(0\le \phi_{p,1}^* \le \phi_u\). The offset angle of the optimal PA angular position relative to the user is denoted as \(\phi_0^*\), and we have 
\begin{align}
\phi_0^* = \phi_u - \phi_{p,1}^*.
\label{eq:phi0_star_first}   
\end{align}

\begin{proposition}
		\label{Proposition1}

The closed-form expression for the optimal PA angle  without blockage \(\phi_{p,1}^*\)  is given by  
\begin{align}
\phi_{p,1}^* = \left[ \phi_u - \arcsin \frac{\rho^2 + r_{wg}^2 + h^2}{2\rho \sqrt{r_{wg}^2 + 1/\alpha^2}} + \arctan(\alpha r_{wg}) \right]^+,
\label{phi_p_star1}
\end{align}
where \( [x]^+ = \max(0, x) \).
	\end{proposition}
 \begin{proof}
		See Appendix A.
	\end{proof}
    
\begin{remark}
As the height of the curved waveguide $h$ or the loss coefficient $\alpha$ increases, $\phi_{p,1}^*$ decreases, i.e., the PA moves toward the start of the waveguide. This indicates that larger $h$ and $\alpha$ lead to a larger deviation of the optimal position from the user's angle $\phi_u$, thereby making the optimal PA placement strategy more advantageous.
\end{remark}
\begin{remark}
In the case that $h$ or $\alpha$ exceeds a certain threshold, the $\arcsin(\cdot)$ term in \eqref{phi_p_star1} has no solution, and thus $\phi_{p,1}^* = 0$. Therefore, the PA is forced to be fixed at the waveguide start and cannot exploit the flexible deployment offered by the C-PAS.
\end{remark}
\subsubsection{Optimal PA Placement with Blockage}

\begin{figure}[t]
	\centering
	\includegraphics[width=\linewidth]{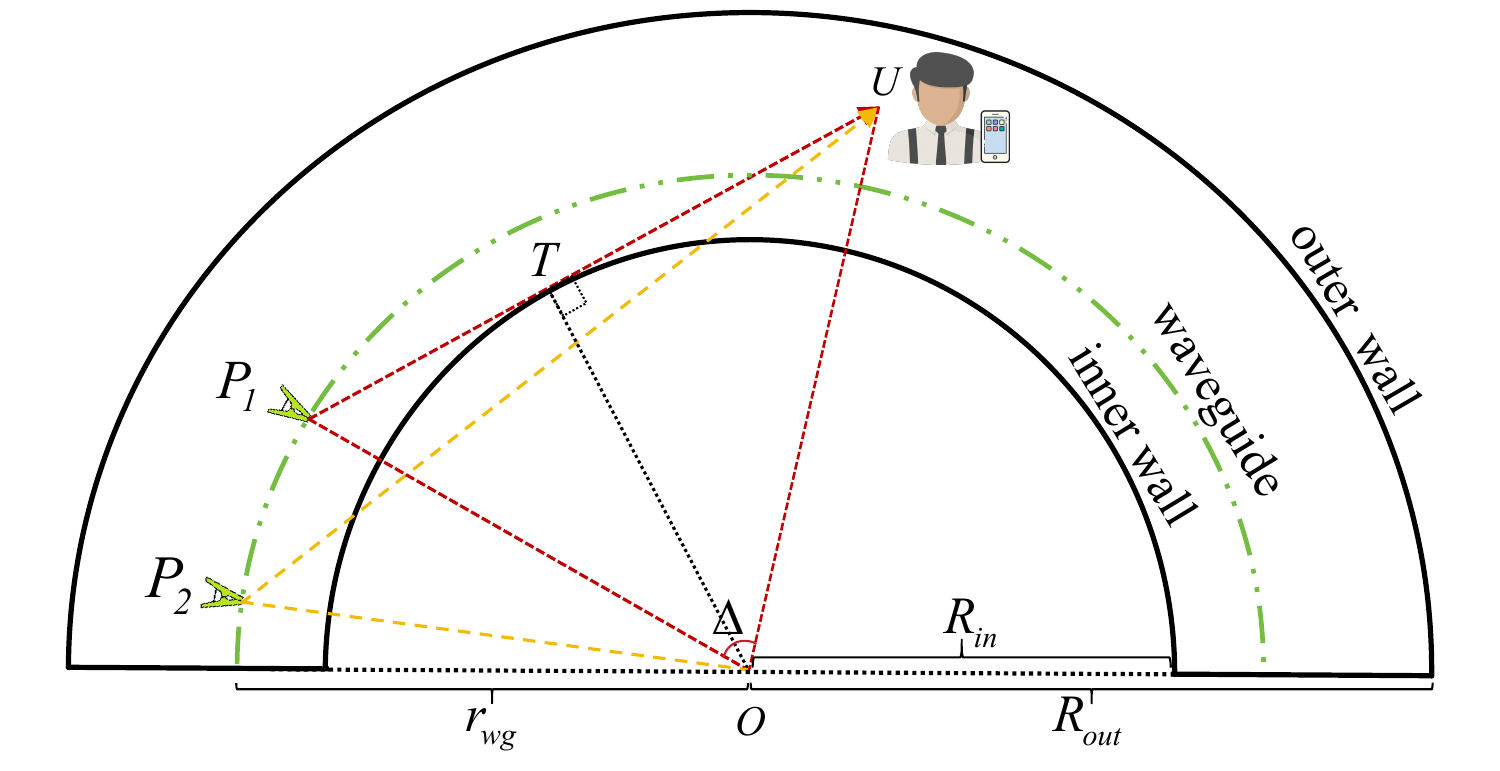}
	\captionsetup{font={footnotesize}} 
	\caption{Planar layout of C-PAS with blockage.}
	\label{fig:blockage}
\end{figure}

As shown in Fig.~\ref{fig:blockage}, the two typical positions of the PA on the same waveguide are denoted as $P_i$ for $i=1,2$, and the position of the user is denoted as $U$. Both lines $UP_1$ (the line connecting points $U$ and $P_1$) and $UP_2$ (the line connecting points $U$ and $P_2$) are blocked by the inner wall, which prevents LoS communication.

Consider the critical case where the line ${UP_1}$ is tangent to the inner wall. From the origin \(O\), draw a perpendicular to the line \({UP_1}\), meeting it at \(T\), i.e., \({OT} \perp {UP_1}\). As the length of ${OT}$ equals \(R_{\text{in}}\), the length of ${OP_1}$ equals \(r_{wg}\), and the length of ${OU}$ equals \(\rho\), we have
\begin{align}
\angle UOT &= \arccos \frac{R_{\text{in}}}{\rho},\\
\angle TOP_1 &= \arccos \frac{R_{\text{in}}}{r_{wg}}.
\end{align}

The angular separation between the PA and the user in the aforementioned critical case corresponds to the LoS occlusion range, denoted by \(\Delta\), is expressed as
\begin{align}
\label{delta_def}
\Delta = \angle UOT + \angle TOP_1 = \arccos \frac{R_{\text{in}}}{\rho} + \arccos \frac{R_{\text{in}}}{r_{wg}}.
\end{align}
\begin{remark}
For a fixed inner-wall radius \(R_{\mathrm{in}}\), the LoS range for a user at \((\rho, \phi_u)\) depends on the waveguide bending radius \(r_{wg}\). In particular, increasing \(r_{wg}\) decreases \(\Delta\) and thus enlarges the LoS range.
\end{remark}

Based on \eqref{delta_def}, we compare \(\angle UOP\), i.e., the subtended angle at \textit{O} between the user and the PA, with \(\Delta\). This comparison decides whether the PA is within LoS range, and the details follow.
\begin{itemize}
\item When \(\angle UOP>\Delta\), the line ${UP}$ intersects the inner wall, indicating that LoS transmission is blocked.
\item When \(\angle UOP = \Delta\), the line ${UP}$ is tangent to the inner wall, indicating the critical condition for LoS transmission.
\item When \(\angle UOP<\Delta\), the line ${UP}$ is separated from the inner wall, indicating that the LoS transmission requirement is satisfied.
\end{itemize}

Since \(\angle UOP = \phi_p - \phi_u\), the mathematical condition for the PA to satisfy LoS communication can be expressed as  
\begin{align}
	\label{los_condition}
	\phi_p > \phi_u - \Delta.
\end{align}

In the case that the PA position in \eqref{phi_p_star1} is blocked by the inner wall, the closed-form expression for the optimal PA angle \(\phi_{p,2}^*\) is given by
\begin{align}
\label{phi_p_star}
\phi_{p,2}^* = \left(\phi_u - \arccos \frac{R_{\mathrm{in}}}{\rho} - \arccos \frac{R_{\mathrm{in}}}{r_{wg}} \right)^+.
\end{align}

\begin{remark}
Taking into account both the above cases with and without LoS blockage, the optimal PA position can be uniformly expressed as
\begin{align}
\label{eq:optimal_phi_p_star}
\phi_p^* = 
\begin{cases}
\left(\phi_u - \phi_0^* \right)^+, & D \le 1 \text{ and } \Delta \ge \phi_0^*, \\[4pt]
\left(\phi_u - \Delta \right)^+, & \text{otherwise},
\end{cases}
\end{align}
where 	$\phi_0^* = \arcsin \frac{\rho^2 + r_{wg}^2 + h^2}{2\rho \sqrt{r_{wg}^2 + 1/\alpha^2}} - \arctan(\alpha r_{wg})$, $\Delta = \arccos \frac{R_{\text{in}}}{\rho} + \arccos \frac{R_{\text{in}}}{r_{wg}}$, and  $D = \frac{\alpha(\rho^2 + r_{wg}^2 + h^2)}{2\rho\sqrt{1+\alpha^2 r_{wg}^2}}$ .
\end{remark}

\subsection{Nearest PA Placement Strategy}
We propose a nearest strategy for the C‑PAS as a benchmark to evaluate the optimal PA placement strategy. Derived from the conventional straight-waveguide approach, this nearest strategy places the PA at the closest position to the user while avoiding inner-wall blockage, which differs from the complex design of the optimal strategy. This strategy is referred to as the \textbf{nearest PA placement strategy}.

 Given that the waveguide bending radius \(r_{wg}\) is fixed, the position of the PA on the waveguide is determined by the angular parameter \( \phi_p \). For a user with coordinates \((\rho, \phi_u)\), the angular position of the PA is given by
  \begin{align}
 \phi_p^{\text{near}} = \phi_u.
 \end{align}
Thus the received SNR of the user, given by \eqref{gamma_r_final}, can be rewritten as
\begin{align}
	\label{gamma_r_arc_N}
	\gamma_r^{\text{near}} = \frac{\eta P_t e^{-\alpha r_{wg} \phi_u} / \sigma^2}{(r_{wg}-\rho)^2 + h^2}.
\end{align}
 \begin{remark}
As shown in \eqref{gamma_r_arc_N}, the SNR received at the user decays exponentially with \(\phi_u\). For users located far from the feed point, the path loss of the curved waveguide dominates, severely degrading the SNR. Therefore, despite minimizing the free-space distance, the nearest strategy still incurs significant SNR loss when the waveguide loss is non-negligible.
\end{remark}
\begin{remark}
As \(r_{wg}\) increases, the received SNR at the user varies nonlinearly. Specifically, for small \(r_{wg}\), the PA is far from the user, resulting in a low SNR. As \(r_{wg}\) further increases, the PA moves closer to the user, reducing the free-space distance and improving the SNR, although the waveguide loss also increases. However, once \(r_{wg}\) exceeds \(\rho\), both the free-space distance and the waveguide loss \(e^{-\alpha r_{wg}\phi_u}\) increase, causing a decrease in SNR.
\end{remark}

\section{Performance Analysis}
This section analyzes the performance of the C-PAS under two placement strategies, and derives the analytical expressions for OP and AR, respectively. The insights are obtained to provide a guideline for the system design.
\subsection{Outage probability}
The OP is defined as the probability that the received SNR $\gamma_r$ does not exceed a given threshold $\gamma_{\text{thr}}$, which is given by
\begin{align}
	\label{P_out_def}
	P_{\mathrm{out}} = \Pr\left(\gamma_r \leq \gamma_{\mathrm{thr}}\right).
\end{align}
Since the user coordinates  $(\rho, \phi_u)$ are uniformly distributed over the arc-shaped region with area $S = \frac{\theta}{2}(R_{\mathrm{out}}^2 - R_{\mathrm{in}}^2)$, the joint probability density function (PDF) can be expressed as
\begin{align}
	{f_{\rho ,{\phi _u}}}(\rho ,{\phi _u}) 
	= \dfrac{1}{S}
	= \dfrac{2}{\theta (R_{\mathrm{out}}^2 - R_{\mathrm{in}}^2)}.
\end{align}
Then the OP over the arc-shaped region is given by
\begin{equation}
\label{eq:outage_integral}
\begin{aligned}
P_{\mathrm{out}} &= \iint_{\gamma_r \leq \gamma_{\mathrm{thr}}} f_{\rho,\phi_u} \, \rho \, d\rho \, d\phi_u \\
&= \frac{2}{\theta \left( R_{\mathrm{out}}^2 - R_{\mathrm{in}}^2 \right)} \int_{R_{\mathrm{in}}}^{R_{\mathrm{out}}} \rho \int_{0}^{\theta} \mathbf{1}_{\{\gamma_r \leq \gamma_{\mathrm{thr}}\}} \, d\phi_u \, d\rho,
\end{aligned}
\end{equation}
where $\mathbf{1}_{\{\cdot\}}$ is the indicator function, which equals $1$ in case  that the condition holds and $0$ otherwise.
By converting the double integral in \eqref{eq:outage_integral} into a single integral over $\rho$, we have
\begin{equation}
\label{eq:outage_radial}
P_{\mathrm{out}} = \frac{2}{\theta \left( R_{\mathrm{out}}^2 - R_{\mathrm{in}}^2 \right)} \int_{R_{\mathrm{in}}}^{R_{\mathrm{out}}} \rho \, L(\rho) \, d\rho,
\end{equation}
where \(L(\rho)\) denotes the angular measure of the outage region at a given radial distance \(\rho\), and is specifically given by
\begin{align}
    L(\rho) = \int_{0}^{\theta} \mathbf{1}_{\{\gamma_r \leq \gamma_{\mathrm{thr}}\}} \, d\phi_u.
    \label{eq:L_total}
\end{align}

\subsubsection{OP for the Optimal PA Placement Strategy}
By evaluating the single integral in \eqref{eq:L_total} and substituting the result into \eqref{eq:outage_radial}, the analytical expression for OP under the optimal PA placement strategy is obtained as follows.
\begin{proposition}
		\label{Proposition2}
The analytical expression of OP under the optimal PA placement strategy is given by
\begin{equation}
\label{eq:outage_indicator}
\begin{aligned}
&P_{\mathrm{out}}^{\text{opt}} = \int_{R_{\mathrm{in}}}^{R_{\mathrm{out}}} \Bigl\{
\mathbf{1}_1 \Bigl[ \theta - \phi_0^* - \frac{\left( \ln\frac{\eta P_t}{\sigma^2\gamma_{\mathrm{thr}}(d_{\max}^2 - 2\rho r_{wg}\cos\phi_0^*)} \right)^+}{\alpha r_{wg}} \Bigr]^+ \\
&+ \mathbf{1}_2 \Bigl[ \min(\phi_0^*,\Delta) - \arccos\!\Bigl(\frac{d_{\max}^2}{2\rho r_{wg}} - \frac{\eta P_t}{2\rho r_{wg}\sigma^2\gamma_{\mathrm{thr}}}\Bigr) \Bigr]^+ \\
&+ \mathbf{1}_3 \Bigl[ \min(\theta,\Delta) - \arccos\!\Bigl(\frac{d_{\max}^2}{2\rho r_{wg}} - \frac{\eta P_t}{2\rho r_{wg}\sigma^2\gamma_{\mathrm{thr}}}\Bigr) \Bigr]^+ \\
&+ \mathbf{1}_4 \Bigl[ \theta - \Delta - \frac{\left( \ln\frac{\eta P_t}{\sigma^2\gamma_{\mathrm{thr}}(d_{\max}^2 - 2\rho r_{wg}\cos\Delta)} \right)^+}{\alpha r_{wg}} \Bigr]^+
\Bigr\} \frac{\rho}{S}\,d\rho,
\end{aligned}
\end{equation}
where $d_{\max}^2 = \rho^2 + r_{wg}^2 + h^2$, $S = \frac{1}{2}\theta (R_{\mathrm{out}}^2 - R_{\mathrm{in}}^2)$, $\mathbf{1}_1=\mathbf{1}_{\substack{D\le1,\ \phi_u>\phi_0^*,\ \Delta>\phi_0^*}}$, $\mathbf{1}_2=\mathbf{1}_{\substack{D\le1,\ \phi_u\le\phi_0^*,\ \Delta>\phi_u}}$, $\mathbf{1}_3=\mathbf{1}_{\substack{D>1,\ \Delta >\phi_u}}$, and  $\mathbf{1}_4=1-\mathbf{1}_1-\mathbf{1}_2-\mathbf{1}_3$.
	\end{proposition}

   \begin{proof}
		See Appendix B.
	\end{proof}

\begin{remark}
In \eqref{eq:outage_indicator}, 
\(\mathbf{1}_1\) corresponds to the case where there is no LoS blockage between the user and the PA, i.e., \(\phi_p^* = \phi_u - \phi_0^*\),
whereas \(\mathbf{1}_4\) corresponds to the case where the PA is placed at the LoS boundary of the inner wall, i.e., \(\phi_p^* = \phi_u - \Delta\), which is the scenario with LoS blockage.
It is worth noting that \(\mathbf{1}_2\) and \(\mathbf{1}_3\) represent the case where \eqref{eq:optimal_phi_p_star} takes the value 0, i.e., \(\phi_p^* = 0\), which corresponds to the cases where the PA is at the starting point of the waveguide, as any movement toward the user would incur more signal attenuation.
\end{remark}

\begin{remark}
As the transmit power \(P_t\) increases, the weights of the individual terms in the outage probability expression \eqref{eq:outage_indicator} change. At high power, the terms associated with \(\mathbf{1}_2\) and \(\mathbf{1}_3\) have no real solution and their contributions become negligible; meanwhile, the logarithmic terms in the \(\mathbf{1}_1\) and \(\mathbf{1}_4\) terms dominate. To minimize the outage probability, we need to minimize the denominator \(\alpha r_{wg}\) corresponding to these logarithmic terms, and thus the smallest  bending  radius of the waveguide \(r_{wg}=R_{\mathrm{in}}\) should be chosen.
\end{remark}

\subsubsection{OP for the Nearest PA Placement Strategy}
By substituting the SNR expression in \eqref{gamma_r_arc_N} into \(\eqref{eq:outage_radial}\) and evaluating the single integral over $\phi_u$, the analytical expression for the OP under the nearest PA placement strategy is obtained as follows.

\begin{proposition}
		\label{Proposition3}
The expression for OP under the nearest PA placement strategy can be expressed as
\begin{align}
\label{outage_closest}
P_{\text{out}}^{\text{near}} = \frac{1}{S} \int_{R_{\text{in}}}^{R_{\text{out}}} \left[ \theta - \frac{1}{\alpha r_{wg}} \ln \frac{\eta P_t / \sigma^2}{\gamma_{\text{thr}} \big[ (r_{wg} - \rho)^2 + h^2 \big]} \right]^+ \rho\,d\rho.
\end{align}
\end{proposition}
   \begin{proof}
With the SNR expression in \eqref{gamma_r_arc_N}, the condition for outage, i.e., $\gamma_r \le \gamma_{\text{thr}}$, can be equivalently written as
\begin{align}
\label{eq:phi_th0}
\phi_u \ge \frac{1}{\alpha r_{wg}} \ln \frac{\eta P_t / \sigma^2}{\gamma_{\mathrm{thr}} [(\rho - r_{wg})^2 + h^2]} \buildrel \Delta \over = \phi_{\mathrm{th}}^0.
\end{align}
For a fixed radial distance \(\rho\), the range of \(\phi_u\) is \([0, \theta]\). Thus the angular interval where outage occurs is \([\phi_{\mathrm{th}}^0, \theta]\), and the interval length is
\begin{align}
\label{eq:L_total1}
L_(\rho) = (\theta - \phi_{\mathrm{th}}^0)^+.
\end{align}
Substituting \(\eqref{eq:L_total1}\) into
\(\eqref{eq:outage_radial}\), the analytical expression for  OP under nearest PA placement strategy is obtained as \eqref{outage_closest}.
	\end{proof}

\begin{remark}
From \eqref{outage_closest}, in the case that the waveguide loss \(\alpha\) or the waveguide height \(h\) is small, the free-space distance term \((r_{wg}-\rho)^2\) dominates. To minimize the outage probability, one should choose \(r_{wg} = \frac{R_{\text{in}}+R_{\text{out}}}{2}\). This middle arc deployment minimizes the average of \((\rho-r_{wg})^2\) over \(\rho \in [R_{\text{in}}, R_{\text{out}}]\), thereby reducing the average free-space attenuation and achieving the lowest outage probability. Meanwhile, under this condition, the offset of the optimal PA position from the user angle is extremely small, so its performance is nearly identical to that of the nearest PA strategy.
\end{remark}

\subsection{Average Rate}
In order to investigate the data transmission capability of the C-PAS, we use the AR as a performance metric.
The AR is the expected value of the achievable Shannon rate.
This expectation is taken over the distribution of user locations and can be expressed as
\begin{align}
	\label{avg_rate_def}
	R_{p} = \mathbb{E}\left[ \log_2 \left(1 + \gamma(\rho, \phi_u)\right) \right].
\end{align}
Expanding the expectation into an integral, the AR of \eqref{avg_rate_def} is given by
\begin{align}
	\label{avg_rate_int}
	R_{p} = \frac{2}{\theta \left( R_{\mathrm{out}}^2 - R_{\mathrm{in}}^2 \right)} \int_{R_{\mathrm{in}}}^{R_{\mathrm{out}}} \rho \left[ \int_0^\theta \log_2(1 + \gamma) \, d\phi_u \right] d\rho.
\end{align}
By converting the double integral in \eqref{avg_rate_int} into a single integral over $\rho$, we have
\begin{align}
	\label{avg_rate_final}
	R_{p}= \frac{2}{\theta \left( R_{\mathrm{out}}^2 - R_{\mathrm{in}}^2 \right)} \int_{R_{\mathrm{in}}}^{R_{\mathrm{out}}} \rho \, J(\rho) \, d\rho,
\end{align}
where \(J_(\rho)\) denotes the rate measure corresponding to all user positions within the angular interval \([0, \theta]\) for a fixed \(\rho\). Then \(J_(\rho)\) is specifically given by
\begin{align}
	\label{J_angle_def}
	J_(\rho) = \int_0^{\theta} \log_2 \left(1 + \gamma\left(\rho, \phi_u\right)\right) d\phi_u.
\end{align}

\subsubsection{AR for the Optimal PA Placement Strategy}
By evaluating the single integral in 
\(\eqref{J_angle_def}\) and substituting the result into 
\(\eqref{avg_rate_final}\), the analytical expression for the AR under the optimal PA placement strategy is obtained as follows.

\begin{figure*}[!t]
\centering
\begin{equation}
\label{eq:R}
\begin{aligned}
R_{p}^{\text{opt}} = &\int_{R_{\mathrm{in}}}^{R_{\mathrm{out}}} \frac{2\rho}{\theta (R_{\mathrm{out}}^2 - R_{\mathrm{in}}^2)} \cdot \Bigl\{ 
\mathbf{1}_A \cdot \frac{1}{\alpha r_{wg} \ln 2}\Biggl[ \operatorname{Li}_2\!\Bigl(-\frac{\frac{\eta P_t}{\sigma^2}}{d_{\max}^2 - 2\rho r_{wg}\cos\phi_0^*}\, e^{-\alpha r_{wg}(\theta - \phi_0^*)}\Bigr) - \operatorname{Li}_2\!\Bigl(-\frac{\frac{\eta P_t}{\sigma^2}}{d_{\max}^2 - 2\rho r_{wg}\cos\phi_0^*}\Bigr) \Biggr] \\
&+ \mathbf{1}_B \cdot \frac{1}{\alpha r_{wg} \ln 2}\Biggl[ \operatorname{Li}_2\!\Bigl(-\frac{\frac{\eta P_t}{\sigma^2}}{d_{\max}^2 - 2\rho r_{wg}\cos\Delta}\, e^{-\alpha r_{wg}(\theta - \Delta)}\Bigr) - \operatorname{Li}_2\!\Bigl(-\frac{\frac{\eta P_t}{\sigma^2}}{d_{\max}^2 - 2\rho r_{wg}\cos\Delta}\Bigr) \Biggr] \\
&+ \mathbf{1}_C \cdot \frac{1}{\ln 2}\Biggl[ \phi_{\mathrm{up}} \ln\frac{\frac{\eta P_t}{\sigma^2} + d_{\max}^2 + \sqrt{\bigl(\frac{\eta P_t}{\sigma^2} + d_{\max}^2\bigr)^2 - (2\rho r_{wg})^2}}{d_{\max}^2 + \sqrt{d_{\max}^4 - (2\rho r_{wg})^2}} \\
&\qquad - 2\operatorname{Im}\Biggl( \operatorname{Li}_2\!\Bigl(\frac{2\rho r_{wg}}{\frac{\eta P_t}{\sigma^2} + d_{\max}^2 + \sqrt{(\frac{\eta P_t}{\sigma^2} + d_{\max}^2)^2 - (2\rho r_{wg})^2}}\, e^{i\phi_{\mathrm{up}}}\Bigr) - \operatorname{Li}_2\!\Bigl(\frac{2\rho r_{wg}}{d_{\max}^2 + \sqrt{d_{\max}^4 - (2\rho r_{wg})^2}}\, e^{i\phi_{\mathrm{up}}}\Bigr) \Biggr) \Biggr]
\Bigr\} d\rho,
\end{aligned}
\end{equation}
\vspace{1ex}
\hrule
\end{figure*}

\begin{proposition}
\label{Proposition4}
The analytical expression for the AR under the optimal PA placement strategy can be expressed as \eqref{eq:R}, shown at the top of the next page, where $\mathbf{1}_A=\mathbf{1}_{\phi_p^* = \phi_u - \phi_0^*}$,
$\mathbf{1}_B = \mathbf{1}_{\phi_p^* = \phi_u - \Delta}$, and $\mathbf{1}_C = \mathbf{1}_{\phi_p^* = 0}$. Moreover, the value of $\phi_{\mathrm{up}}$ depends on the parameter conditions. Specifically, when $D \le 1$ and $\phi_u \le \phi_0^*$, then $\phi_{\mathrm{up}} = \min(\phi_0^*, \Delta)$; whereas when $D > 1$ and $\Delta \ge \phi_u$, then $\phi_{\mathrm{up}} = \Delta$.
\end{proposition}

	\begin{proof}
		See Appendix C.
	\end{proof}

\begin{remark}
In \eqref{eq:R}, \(\mathbf{1}_A\) corresponds to the case where there is no LoS blockage between the user and the PA, and the PA is not at the waveguide origin, i.e., \(\phi_p^* = \phi_u - \phi_0^*\). 
\(\mathbf{1}_B\) corresponds to the scenario where the PA is forced to the LoS boundary of the inner wall due to blockage, i.e., \(\phi_p^* = \phi_u - \Delta\). 
\(\mathbf{1}_C\) corresponds to the scenario where the PA is fixed at the starting point of the waveguide, i.e., \(\phi_p^* = 0\).
\end{remark}

\subsubsection{AR for the Nearest PA Placement Strategy}            
Similarly, substituting the nearest PA SNR from \eqref{gamma_r_arc_N} into \eqref{avg_rate_final} and performing the integration over the angular domain yields the following analytical expression for the AR.

\begin{proposition}
 The AR under the nearest PA deployment can be expressed as
\begin{align}
R_{p}^{\text{near}}= & \int_{R_{\mathrm{in}}}^{R_{\mathrm{out}}} \frac{\rho}{S \, \alpha r_{wg} \ln 2} \Bigg[ 
\operatorname{Li}_2\!\left(-\frac{\eta P_t \, e^{-\alpha r_{wg} \theta} / \sigma^2}{(\rho - r_{wg})^2 + h^2}\right) \nonumber \\
&- \operatorname{Li}_2\!\left(-\frac{\eta P_t / \sigma^2}{(\rho - r_{wg})^2 + h^2}\right) \Bigg] d\rho.
\label{eq:r_{wg}_nearest}
\end{align}
 \end{proposition}
 
   \begin{proof}
		Substituting \eqref{gamma_r_arc_N} into \eqref{J_angle_def}, we have
\begin{align}
J_(\rho) &= \int_{0}^{\theta} \log_2\!\left(1 + \frac{\eta P_t \, e^{-\alpha r_{wg} \phi_u}}{\sigma^2 \left[(\rho - r_{wg})^2 + h^2\right]}\right) d\phi_u \nonumber \\
&= \int_{0}^{\theta} \frac{\ln\!\left(1 + \frac{\eta P_t}{\sigma^2 \left[(\rho - r_{wg})^2 + h^2\right]} \cdot e^{-\alpha r_{wg} \phi_u}\right)}{\ln 2} \, d\phi_u.
\end{align}
By the substitution \(t = e^{-\alpha r_{wg} \phi_u}\), we have
\begin{align}
J_(\rho) &= \frac{1}{\alpha r_{wg} \ln 2} \int_{1}^{e^{-\alpha r_{wg} \theta}} \frac{\ln\!\left(1 + \frac{\eta P_t}{\sigma^2 \left[(\rho - r_{wg})^2 + h^2\right]} \cdot t\right)}{t} \,dt .
\label{eq:J_rho}
\end{align}
Using the definition of the dilogarithm function in \cite[Eq. (6.254.1)]{gradshteyn2007table}, \eqref{eq:J_rho} can be written as
\begin{align}
J_(\rho) = & \frac{1}{\alpha r_{wg} \ln 2} \Bigg[ 
\operatorname{Li}_2\!\left(-\frac{\eta P_t \, e^{-\alpha r_{wg} \theta} / \sigma^2}{(\rho - r_{wg})^2 + h^2}\right) \nonumber \\
& - \operatorname{Li}_2\!\left(-\frac{\eta P_t / \sigma^2}{(\rho - r_{wg})^2 + h^2}\right) \Bigg].
\label{eq:J_rho_solved}
\end{align}
Substituting \eqref{eq:J_rho_solved} into \eqref{avg_rate_final}, the analytical expression for AR under nearest PA placement strategy is obtained as \eqref{eq:r_{wg}_nearest}.
	\end{proof}

\begin{remark}
For fixed \(S\) and \(R_{\text{in}}\), an increase in \(\theta\) reduces \(R_{\text{out}}=\sqrt{R_{\text{in}}^2+2S/\theta}\), thereby shortening the integration interval. At the same time, the exponential factor \(e^{-\alpha r_{wg}\theta}\) decreases, which reduces the absolute value of the argument of the first dilogarithm term. Since \(\operatorname{Li}_2(-x)\) is monotonically decreasing and negative for \(x>0\), this term increases. The competition between these two opposing effects makes \(R_{p}^{\text{near}}(\theta)\) first increase and then decrease, so that there exists a unique \(\theta\) that maximizes the average rate.
\end{remark}

\begin{remark}
Under the constraint of fixed area \(S\) and fixed angle \(\theta\), as \(R_{\text{in}}\) increases, the integration interval \([R_{\text{in}}, R_{\text{out}}]\) shrinks monotonically in width. Meanwhile, the average distance from users to the waveguide decreases, i.e., \((\rho-r_{wg})^2\) becomes smaller on average, which increases the difference between the two dilogarithm terms in the integrand. However, the shrinking interval also reduces the total integral value. The competition between these two opposing effects causes \(R_{p}^{\text{near}}\) to first increase and then decrease with \(R_{\text{in}}\), exhibiting a non-monotonic behavior. Consequently, there exists a unique optimal inner-wall radius \(R_{\text{in}}\) that maximizes the average rate.
\end{remark}

\section{Simulation Results and Discussion}
This section presents the performance simulation and analysis of the proposed C-PAS. Without loss of generality, the transmit power of the base station is $P_t = 25\,\text{dBm}$, the height of the curved waveguide is fixed at $h = 3\,\text{m}$, the carrier frequency is set to $f_c = 60\,\text{GHz}$, the noise power is $\sigma^2 = -90\,\text{dBm}$, and the SNR threshold is $\mathrm{SNR}_{\mathrm{thr}} = 20\,\text{dB}$. Unless otherwise specified, the communication service region is set to be an arc-shaped area with inner-wall radius $R_{\text{in}} = 10\,\text{m}$, outer-wall radius $R_{\text{out}} = 20\,\text{m}$, and angle $\theta = 0.5\pi$, respectively. Moreover, the waveguide loss coefficient is $\alpha = 0.1\,\text{Np/m}$\footnote{It is worth noting that bending radiation loss significantly affects system performance only when the bending radius of waveguides is sufficiently small (e.g., below the decimeter scale). In typical indoor ceiling-mounting scenarios, where the bending radius of  waveguides is often on the order of meters, $\alpha_{\text{rad}}$ is extremely small and can be safely neglected. Therefore, the loss coefficient $\alpha$ used in our simulations mainly accounts for material absorption.}. All Monte Carlo simulation results are obtained by averaging over $10^5$ independent channel realizations.

\begin{figure}[t]
	\centering
	\includegraphics[width=\linewidth]{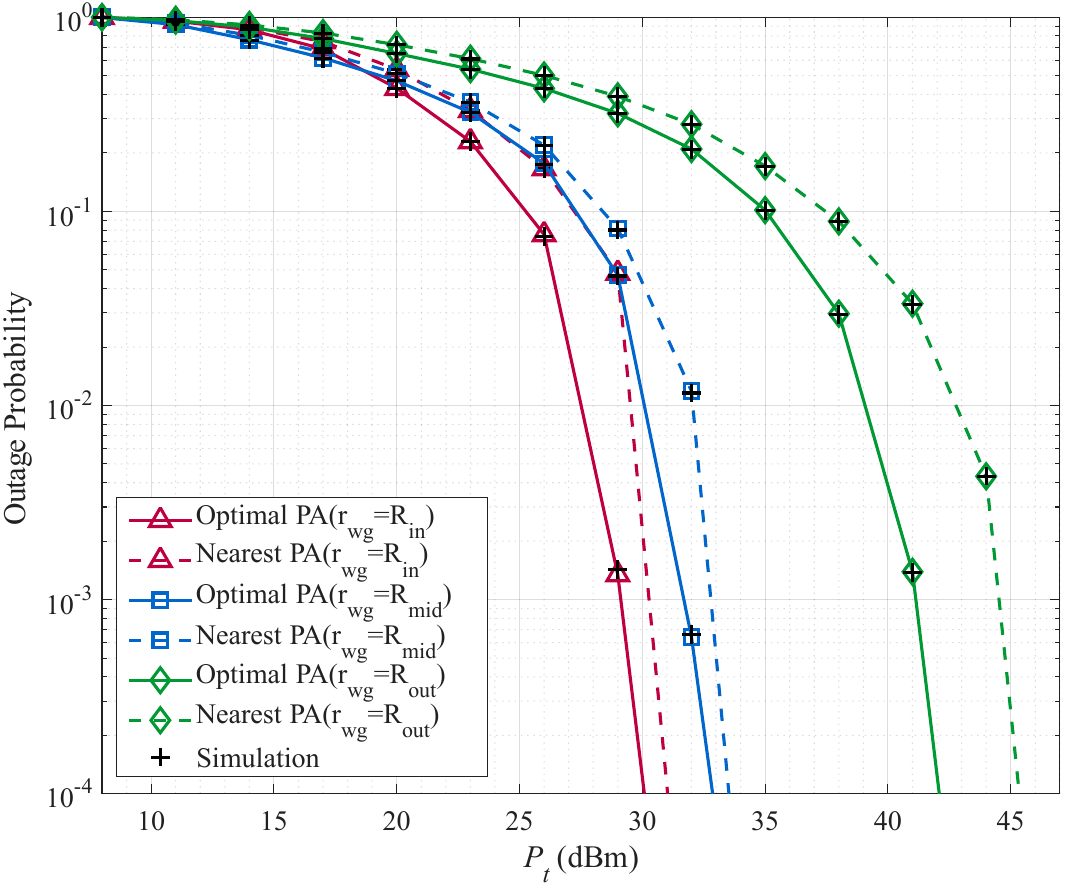}
	\captionsetup{font={footnotesize}} 
	\caption{OP versus \(P_t\) for different strategies with \(\alpha = 0.2\ \text{Np/m}\).}
	\label{P_t-OP}
\end{figure}

Fig.~\ref{P_t-OP} shows the OP versus the transmit power \(P_t\) for different strategies with \(\alpha = 0.2\ \text{Np/m}\).  
It is observed that the analysis expressions of OP match well with the corresponding simulation results, verifying the accuracy of our theoretical analysis.  
The OP for both strategies decreases monotonically with the increase of $P_t$, as a higher $P_t$ improves the SNR received at the user. Note that the OP of the optimal PA placement strategy is consistently lower than that of the nearest PA strategy, which verifies the superiority of the proposed optimal PA placement strategy. 
In addition, the impact of the waveguide bending radius  \(r_{wg}\) on the OP is related to the transmit power. At low \(P_t\), free-space attenuation dominates, and the middle-arc radius \(r_{wg}=R_{\text{mid}}\) performs best as it minimizes the average distance between the user and the curved waveguide. At high \(P_t\), waveguide loss becomes the main limiting factor, and the inner-wall radius \(r_{wg}=R_{\text{in}}\) is superior as it corresponds to the shortest waveguide length.

\begin{figure}[t]
	\centering
	\includegraphics[width=\linewidth]{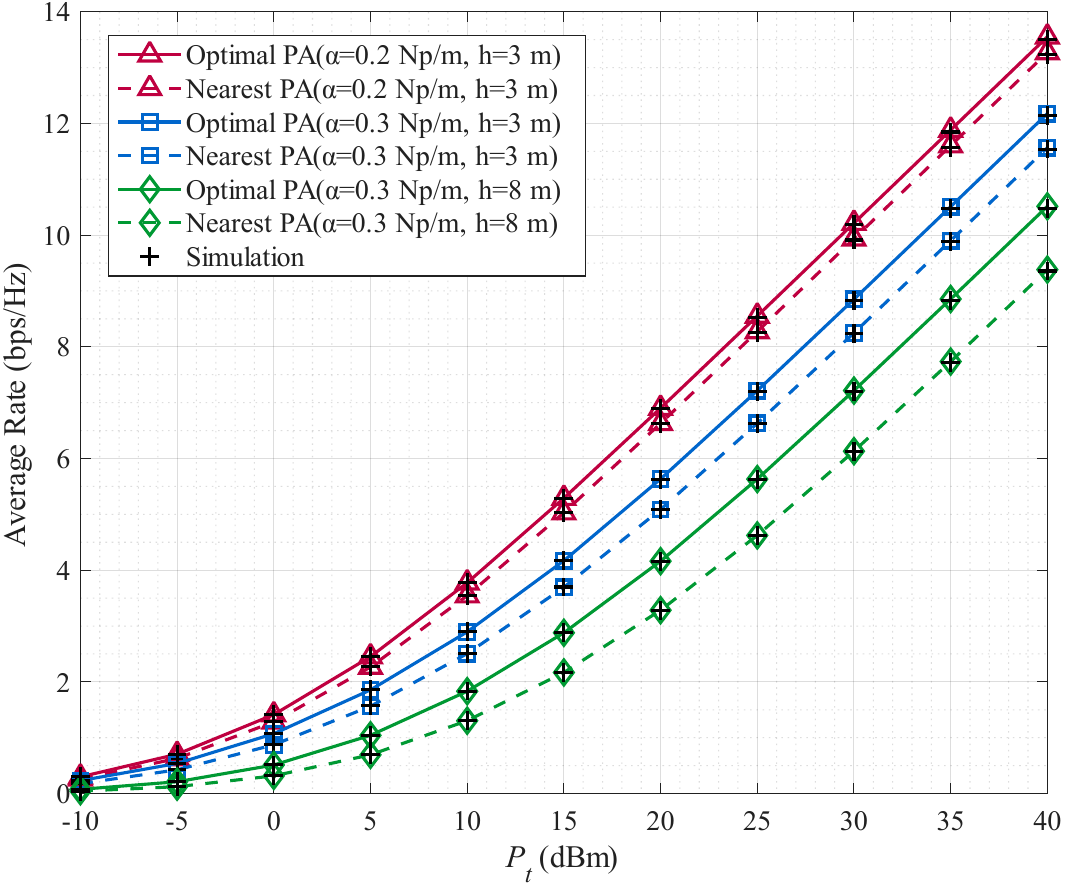}
	\captionsetup{font={footnotesize}} 
	\caption{AR versus \(P_t\) for different strategies  with \(\alpha = 0.2\ \text{Np/m}\) and \(r_{wg}=R_{\text{mid}}\).}
	\label{P_t-R}
\end{figure}
Fig.~\ref{P_t-R} shows the AR versus the transmit power \(P_t\) for different strategies, with \(\alpha = 0.2\ \text{Np/m}\) and \(r_{wg}=R_{\text{mid}}\). As can be seen, 
the analysis expressions of AR match well with the corresponding simulation results, corroborating the accuracy of our theoretical analysis.
It can be seen that the AR of both strategies increases monotonically with \(P_t\).
Notably that the AR gap between the two strategies becomes larger as the loss coefficient \(\alpha\) or the curved waveguide height \(h\) increases. This is because, as \(\alpha\) increases, the waveguide loss becomes more severe, and the nearest strategy cannot effectively cope with it, while the optimal strategy can mitigate the degradation of waveguide channel by adjusting the PA position. As \(h\) increases, the free-space propagation distance increases, causing a severe performance degradation of the nearest strategy; in contrast, the optimal strategy actively moves the PA toward the start of the waveguide, which trades a small increase in free-space loss for a substantial reduction in waveguide loss, thereby maintaining a high received SNR. The above phenomena all indicate that the optimal PA placement strategy exhibits stronger robustness in dealing with channel degradation.

\begin{figure}[t]
	\centering
	\includegraphics[width=\linewidth]{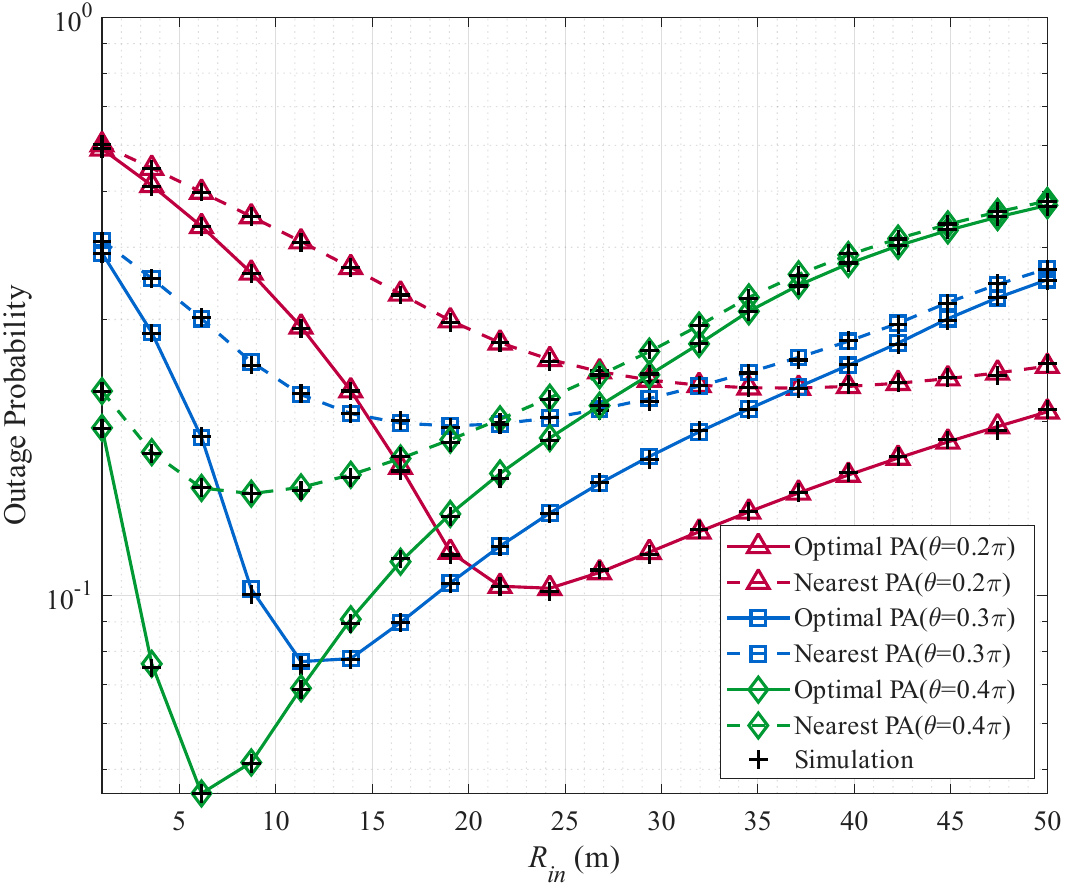}
	\captionsetup{font={footnotesize}} 
	\caption{OP versus \(R_{\text{in}}\) for different strategies with $S = 130\pi\,\text{m}^2$ and \(r_{wg}=R_{\text{in}}\).}
	\label{Rin-OP}
\end{figure}

Fig.~\ref{Rin-OP} shows the OP versus the inner-wall radius \(R_{\text{in}}\) for different strategies under fixed service area $S = 130\pi\,\text{m}^2$ and fixed sector angle \(\theta\), with \(r_{wg}=R_{\text{in}}\).
It can be observed that as \(R_{\text{in}}\) increases, the OP first decreases and then increases. 
\(R_{\text{in}}\) is initially small, and the service area width \(w_0\) is large, resulting in a long average free-space distance between the user and the curved waveguide, which leads to a low received SNR. 
As \(R_{\text{in}}\) increases, \(w_0\) decreases, shortening the average free-space distance thus gradually improving the received SNR.
However, when \(R_{\text{in}}\) continues to increase, the variation in free-space distance becomes slight, but the waveguide length increases rapidly, causing a sharp rise in waveguide loss and simultaneously reducing the LoS  range. At this point, further increasing \(R_{\text{in}}\) leads to a decrease in the received SNR.
Therefore, there exists an optimal \(R_{\text{in}}\) that achieves the best OP performance.
Moreover, there is an interplay between \(R_{\text{in}}\) and \(\theta\) on the OP. For a small \(R_{\text{in}}\), a larger \(\theta\) is superior, whereas for a large \(R_{\text{in}}\), a smaller \(\theta\) is  preferable. The reason is that extreme values of either \(R_{\text{in}}\) or \(\theta\) result in an overly narrow region, causing performance degradation, while moderate values of both parameters enable the OP to reach a relatively ideal level.

\begin{figure}[t]
	\centering
	\includegraphics[width=\linewidth]{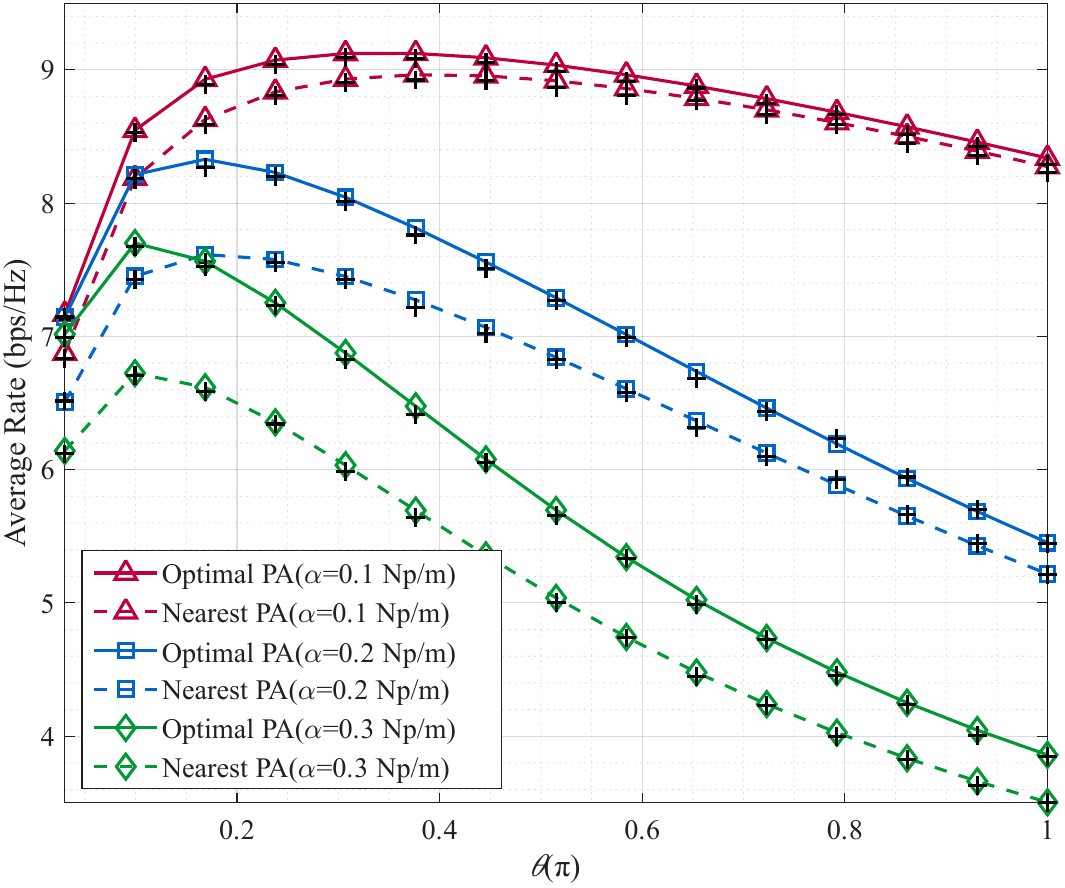}
	\captionsetup{font={footnotesize}} 
	\caption{AR versus $\theta$ for different strategies with $S = 150\pi\,\text{m}^2$ and \(r_{wg}=R_{\text{mid}}\).}
	\label{theta-R}
\end{figure}

Fig.~\ref{theta-R} shows the AR versus the sector angle \(\theta\) under fixed service area $S = 150\pi\,\text{m}^2$ and inner-wall radius \(R_{\text{in}}\), with \(r_{wg}=R_{\text{mid}}\).
It can be observed that as \(\theta\) increases, the AR first increases and then decreases. 
As \(\theta\) is small, the outer-wall radius \(R_{\text{out}}\) is large, resulting in a long free-space distance between the user and the curved waveguide. 
This leads to severe wireless attenuation and a low received SNR. 
As \(\theta\) increases, \(R_{\text{out}}\) decreases, shortening the free-space distance and gradually improving the received SNR. 
However, as \(\theta\) continues to grow to a large value, the variation in free-space distance becomes less pronounced, while the waveguide length increases rapidly, causing a significant rise in waveguide loss and eventually degrading the received SNR.
Furthermore, the optimal \(\theta\) is related to \(\alpha\). For small \(\alpha\), a larger \(\theta\) is preferable, since the waveguide loss is negligible, thereby leading to a more compact service region and reducing the free-space path loss. Conversely, for large \(\alpha\), a smaller \(\theta\) is better, as the waveguide loss becomes the dominant factor limiting the performance, and a smaller \(\theta\) helps avoid excessive waveguide length.

\begin{figure}[t]
	\centering
	\includegraphics[width=\linewidth]{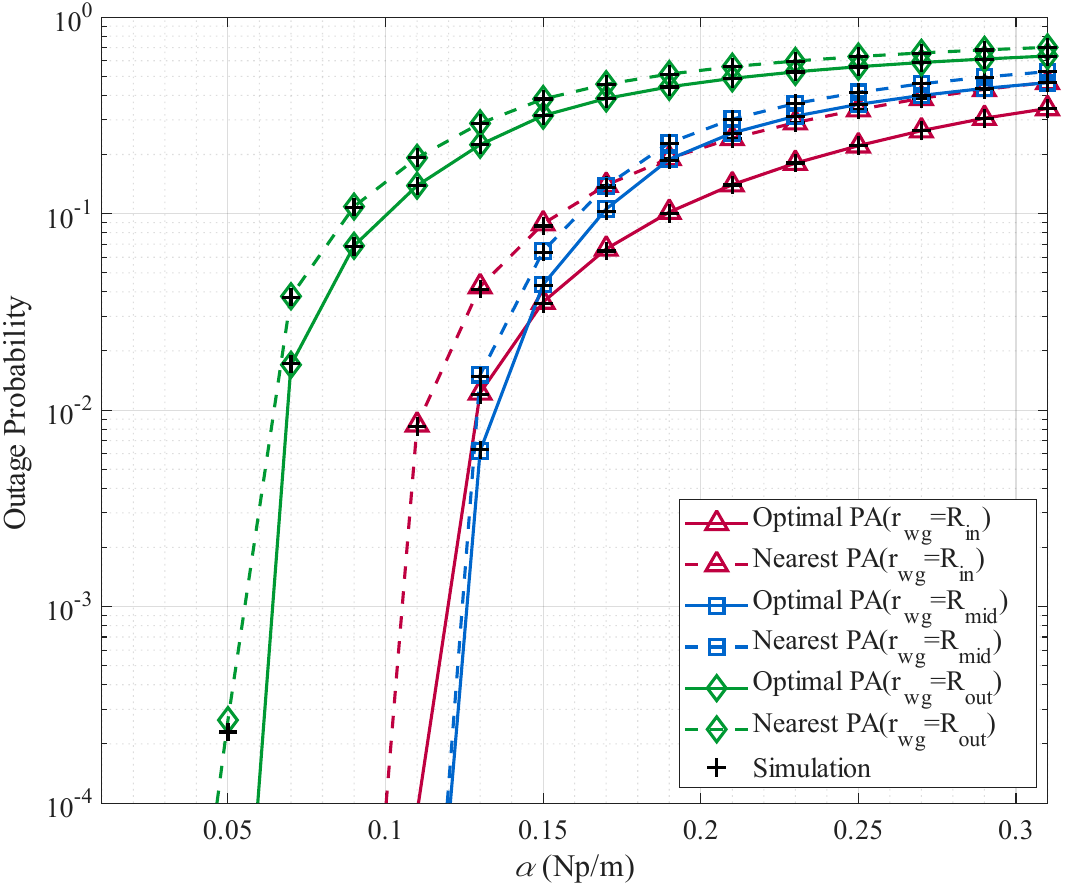}
	\captionsetup{font={footnotesize}} 
	\caption{OP versus \(\alpha\) for different strategies.}
	\label{alpha-OP}
\end{figure}

Fig.~\ref{alpha-OP} shows the OP versus the waveguide loss coefficient \(\alpha\) for different strategies. It can be observed that as \(\alpha\) increases, the OP of both strategies increases monotonically.
The three waveguide bending radii are compared in Fig.~\ref{alpha-OP}. For small \(\alpha\), the OP is lowest with \(r_{wg}=R_{\text{mid}}\). This is because the waveguide loss is negligible and the OP performance is dominated by free-space path loss; the radius \(R_{\text{mid}}\) minimizes the average distance between the user and the curved waveguide, thereby maximizing the received SNR.
For large \(\alpha\), the OP is lowest at \(r_{wg}=R_{\text{in}}\). This is because the waveguide loss becomes severe and there is no theoretically optimal PA position without blockage along the waveguide. Therefore, the PA communicating with the user is often forced to the LoS boundary, and the radius \(R_{\text{in}}\) corresponds to the shortest waveguide length, which minimizes the waveguide loss and  yields the best OP performance.

\begin{figure}[t]
	\centering
	\includegraphics[width=\linewidth]{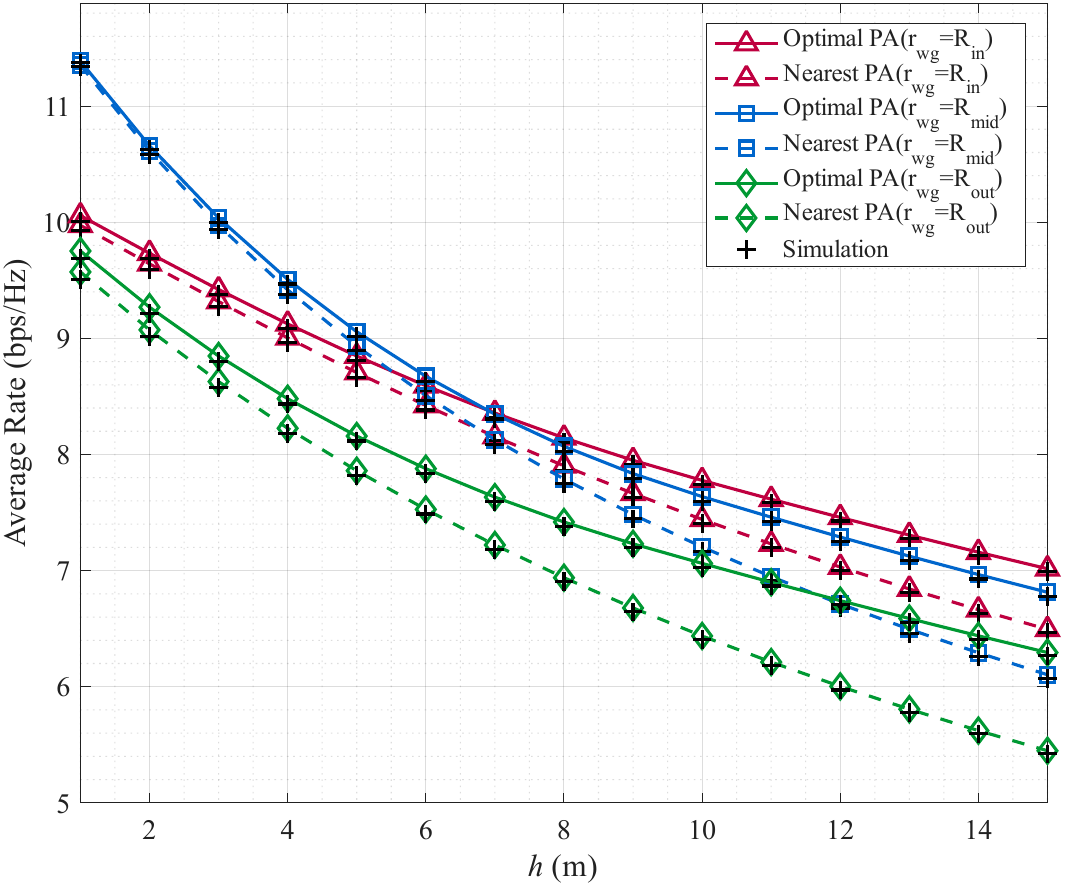}
	\captionsetup{font={footnotesize}} 
	\caption{AR versus $h$ for different strategies.}
	\label{h-R}
\end{figure}

Fig.~\ref{h-R} shows the AR versus the height of the curved waveguide \(h\) for different strategies. It can be observed that as \(h\) increases, the AR of both strategies decreases monotonically, since the free-space propagation distance between the user and the PA increases, leading to a degradation in received SNR.
Furthermore, the choice of the bending radius \(r_{wg}\) of the curved waveguide also varies for different values of \(h\).
For small \(h\), \(r_{wg}=R_{\text{mid}}\) performs best, as the free-space distance is primarily dominated by the horizontal separation; the mid-radius deployment minimizes the average horizontal distance between the user and the waveguide, thus achieving the highest received SNR. For large \(h\), \(r_{wg}=R_{\text{in}}\) performs best, since the vertical distance dominates the wireless propagation loss and weakens the impact of horizontal distance differences among different bending radii. In this case, the waveguide loss becomes the decisive factor limiting the AR. Since \(r_{wg}=R_{\text{in}}\) corresponds to the shortest waveguide length, it incurs the lowest waveguide loss and consequently achieves the highest received SNR.
\begin{figure}[t]
	\centering
	\includegraphics[width=\linewidth]{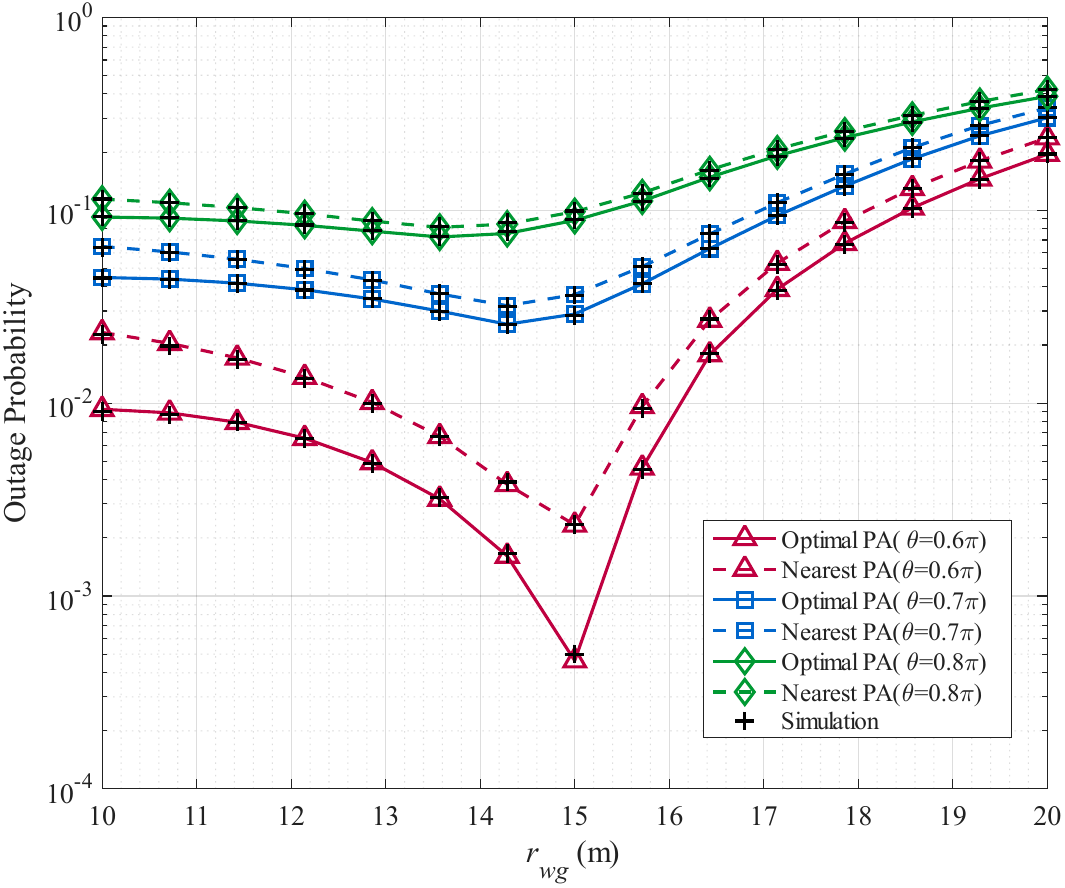}
	\captionsetup{font={footnotesize}} 
	\caption{OP versus \(r_{\text{wg}}\) for different strategies.}
	\label{r-OP}
\end{figure}

Fig.~\ref{r-OP} shows the OP versus the bending radius  \(r_{\text{wg}}\) of the curved waveguide for different strategies. 
It can be observed that the OP first decreases and then increases as \(r_{\text{wg}}\) grows.
 This  is because, for small \(r_{\text{wg}}\),  the average free-space distance to the users is large, leading to a low received SNR.
 As \(r_{\text{wg}}\) moves toward the middle arc \(R_{\text{mid}}\), the average distance to the users decreases, which alleviates free-space fading and also enlarges the LoS coverage. 
Particularly for the optimal PA strategy, it can further adjust the PA position to mitigate waveguide loss, thereby improving the received SNR. 
However, as \(r_{\text{wg}}\) approaches the outer-wall radius \(R_{\text{out}}\), the waveguide length increases, introducing severe waveguide loss, which again degrades the received SNR.
It is noteworthy that the optimal bending radius of the curved waveguide is not fixed at the middle arc \(R_{\text{mid}}\), but shifts toward the inner-wall arc  \(R_{\text{in}}\) as the sector angle \(\theta\) increases. The reason is that \(\theta\) determines the sensitivity of the total waveguide length to \(r_{\text{wg}}\). For large \(\theta\), a slight increase in \(r_{\text{wg}}\) leads to a significant increase in waveguide loss; thus the system tends to choose a smaller \(r_{\text{wg}}\) to mitigate the rapid increase of waveguide loss. Conversely, for small \(\theta\), the waveguide loss increases gently, allowing the system to adopt \(r_{wg}=R_{\text{mid}}\) to reduce free-space path loss. These results indicate that the optimal bending radius of the curved waveguide should be designed according to \(\theta\).

\section{Conclusions}
This paper proposes the concept of C‑PAS and establishes a communication model for the arc‑shaped region that accounts for waveguide loss and inner‑wall obstruction. To the best of our knowledge, this is the first work on PAS with curved waveguides.  Moreover, two PA placement strategies are proposed, i.e., the optimal and the nearest. Subsequently, analytical OP and AR expressions are obtained for both strategies, and the trade‑off among waveguide loss, free‑space attenuation, and LoS blockage is discussed. Simulations validate the theoretical analysis, and the following conclusions are drawn.
1)  The OP and AR of the optimal PA placement strategy are superior to those of the nearest placement strategy, especially for large waveguide loss coefficient or waveguide height.
2) For a service region of fixed area, there exists an optimal sector angle or inner-wall radius that either minimizes the outage probability or maximizes the average rate.
3) At high BS transmit power, a smaller bending radius of the curved waveguide is preferable, while at low power,  the middle arc achieves the best performance.
4) For small waveguide loss coefficient or low waveguide height, the waveguide bending radius should be chosen as the middle arc.
5) For high waveguide height  or large loss coefficient, a smaller bending radius of the curved waveguide  is preferable.

\appendices
\section{PROOF OF PROPOSITION 1 }

\renewcommand{\theequation}{A.\arabic{equation}}
\setcounter{equation}{0} 

Substituting $\phi_0 = \phi_u - \phi_p$ into \eqref{gamma_r_arc}, the SNR can be rewritten as
\begin{equation}
\gamma_r^{\text{opt}} = \frac{\eta P_t e^{-\alpha \phi_u r_{wg} + \alpha \phi_0 r_{wg}}}{\sigma^2 \left( \rho^2 + r_{wg}^2 + h^2 - 2\rho r_{wg} \cos(\phi_0) \right)}.
\label{eq:gamma_r}
\end{equation}
The value of $\phi_0$ that maximizes \eqref{eq:gamma_r}  is equivalent to the value of $\phi_0$ that maximizes $\Omega\left( \phi_0 \right) = \frac{e^{\alpha \phi_0 r_{wg}}}{\rho^2 + r_{wg}^2 + h^2 - 2\rho r_{wg} \cos\left( \phi_0 \right)}$.
Taking the logarithm of $\Omega\left( \phi_0 \right)$, we have
\begin{equation}
\ln \left( {\Omega(\phi_0)} \right) = \alpha \phi _0{r_{wg}} - \ln \left[ {{\rho ^2} + r_{wg}^2 + {h^2} - 2\rho {r_{wg}}\cos (\phi _0)} \right].
\label{eq:ln_phi0}
\end{equation}
Taking the derivative of \eqref{eq:ln_phi0} with respect to $\phi_0$, we have
\begin{equation}
\Omega'(\phi_0) = \Omega(\phi_0) \left[ \alpha r_{wg} - \frac{2\rho r_{wg} \sin(\phi_0)}{\rho^2 + r_{wg}^2 + h^2 - 2\rho r_{wg} \cos(\phi_0)} \right].
\label{eq:Omega_prime}
\end{equation}
When $\Omega'(\phi_0)=0$, we obtain the equation concerning $\phi_0$ is given by
\begin{equation}
\frac{2\rho \sin(\phi_0)}{\alpha} = \rho^2 + r_{wg}^2 + h^2 - 2\rho r_{wg} \cos(\phi_0).
\label{eq:phi0_eq}
\end{equation}
Using the auxiliary angle formula to transform \eqref{eq:phi0_eq}, we have
\begin{equation}
\sin ( \phi_0 + \psi ) = D,
\label{eq:sin_phi0_arctan}
\end{equation}
where $D = \frac{\alpha(\rho^2 + r_{wg}^2 + h^2)}{2\rho\sqrt{1+\alpha^2 r_{wg}^2}}$, and  $\psi = \arctan(\alpha r_{wg})$.

Considering that the principal value range of $\arctan$ is $[-\pi/2, \pi/2]$, the stationary points of \eqref{eq:gamma_r} are obtained by transforming \eqref{eq:sin_phi0_arctan} as 
$\phi_0^{(1)} = \arcsin(D) - \psi$ and$ \quad 
\phi_0^{(2)} = \pi - \arcsin(D) - \psi.$
 Differentiating $\Omega'(\phi_0)$ in \eqref{eq:Omega_prime} again with respect to $\phi_0$, the second-order derivative expression of $\Omega(\phi_0)$ is given by
\begin{equation}
\Omega''(\phi_0) = -2\rho r_{wg} e^{\alpha r_{wg} \phi_0} \cdot \frac{(\rho^2 + r_{wg}^2 + h^2)\cos\phi_0 - 2\rho r_{wg}}{\bigl(\rho^2 + r_{wg}^2 + h^2 - 2\rho r_{wg} \cos\phi_0\bigr)^3}.
\end{equation}
The denominator of \(\Omega''(\phi_0)\) is always positive, so the sign of \(\Omega''(\phi_0)\) is determined by the numerator $M(\phi_0) = (\rho^2 + r_{wg}^2 + h^2)\cos\phi_0- 2\rho r_{wg}$. Substituting \eqref{eq:phi0_eq} into \(M(\phi_0)\), we have
\begin{equation}
\begin{aligned}[b]
M(\phi_0) &= \left( \frac{2\rho \sin\phi_0}{\alpha} + 2\rho r_{wg} \cos\phi_0 \right) \cos\phi_0 - 2\rho r_{wg}\\
&= \frac{2\rho \sin\phi_0 \cos\phi_0}{\alpha} + 2\rho r_{wg} \cos^2\phi_0 - 2\rho r_{wg}\\
&= 2\rho \sin\phi_0 \left( \frac{\cos\phi_0}{\alpha} - r_{wg} \sin\phi_0 \right).
\end{aligned}
\end{equation}
Given that $\sin\phi_0 \geq 0$ for $\phi_0 \in [0,\pi)$, the sign of $M(\phi_0)$ is hence determined by $K(\phi_0)$, which is expressed as 
\begin{align}
K(\phi_0) &= \frac{\cos\phi_0}{\alpha} - r_{wg} \sin\phi_0 \nonumber\\
&= \frac{1}{\alpha}\bigl( \cos\phi_0 - \alpha r_{wg} \sin\phi_0 \bigr) \nonumber\\
&= \frac{\sqrt{1+(\alpha r_{wg})^2}}{\alpha} \cos(\phi_0 + \psi).
\end{align}

Since \(\frac{\sqrt{1+(\alpha r_{wg})^2}}{\alpha} > 0\), the sign of \(K(\phi_0)\) follows that of \(\cos(\phi_0 + \psi)\). According to  \(\sin(\phi_0 + \psi) = D\) in \eqref{eq:sin_phi0_arctan}, we have 

\begin{itemize}
    \item \(\phi_0^{(1)} + \psi = \arcsin D \in [0, \pi/2]\). Then \(\cos(\phi_0^{(1)} + \psi) \ge 0\), so \(K > 0\), which implies \(M > 0\) and consequently \(\Omega'' < 0\). Hence \(\phi_0^{(1)}\) is a maximum point.
    \item \(\phi_0^{(2)} + \psi = \pi - \arcsin D \in [\pi/2, \pi]\). Then \(\cos(\phi_0^{(2)} + \psi) \le 0\), so \(K < 0\), which implies \(M < 0\) and consequently \(\Omega'' > 0\). Hence \(\phi_0^{(2)}\) is a minimum point.
\end{itemize}

Then substitute $\phi_0^{(1)}$  for $\phi^{*}_{0}$ in \eqref{eq:phi0_star_first} to obtain the expression for $\phi^{*}_{p}$.

The proof is completed.

\section{PROOF OF PROPOSITION 2 }
\renewcommand{\theequation}{B.\arabic{equation}}
\setcounter{equation}{0} %
Referring to the piecewise expression of the optimal PA position in \eqref{eq:optimal_phi_p_star}, the following three cases need to be considered when calculating \(L_(\rho)\) in \eqref{eq:L_total}.

\textbf 1) Optimal PA position without Blockage

From \eqref{eq:phi0_star_first}, we have $\phi_{p,1}^* = \phi_u - \phi_0^*$, and substituting it into \eqref{gamma_r_arc}, the SNR can be rewritten as
\begin{equation}
\gamma_r^{\text{opt}}= \frac{A \exp\left[-\beta\left(\phi_u - \phi_0^*\right)\right]}{d_0^2},
\label{eq:B1}
\end{equation}
where $A = \frac{\eta P_t}{\sigma^2}$, $B = 2\rho r_{wg}$, $\beta = \alpha r_{wg}$, $d_{\max}^2 = \rho^2 + r_{wg}^2 + h^2$, and $d_0^2 = d_{\max}^2 - B \cos\phi_0^*$.  
Then the outage condition \(\gamma_r \leq \gamma_{\text{thr}}\) can be expressed as 
\begin{equation}
\frac{A \exp\left[-\beta\left(\phi_u - \phi_0^*\right)\right]}{d_0^2} \leq \gamma_{\text{thr}} .
\label{eq:ineq_phi}
\end{equation}
After taking the natural logarithm of both sides of \eqref{eq:ineq_phi}, we have
\begin{equation}
-\beta\left(\phi_u - \phi_0^*\right) \leq \ln\left( \frac{\gamma_{\text{thr}} d_0^2}{A} \right).
\end{equation}
Accordingly, the outage condition can be expressed as an expression related to $\phi_u$, given by
\begin{equation}
\phi_u \geq \phi_0^* + \frac{1}{\beta} \ln\left( \frac{A}{\gamma_{\text{thr}} d_0^2} \right) = \phi_{\text{th}}^{(A)}.
\end{equation}
The total outage angle is the part within the angular domain $[0, \theta]$ that is not less than $\phi_{\text{th}}^{(A)}$, expressed as 
\begin{equation}
L^{(A)}(\rho) = \left[ \theta - \phi_0^* - \frac{1}{\beta}\max\!\left(0,\; \ln\frac{A}{\gamma_{\text{thr}} d_0^2}\right) \right]^+.
\end{equation}

\textbf 2) Optimal PA position with Blockage

Based on \eqref{phi_p_star}, substituting $\phi_p^* = \phi_u - \Delta$ into \eqref{gamma_r_arc}, the SNR can be rewritten as
\begin{equation}
\gamma_r^{\text{opt}}= \frac{A \exp\left[ -\beta \left( \phi_u - \Delta \right) \right]}{d_1^2},
\label{eq:gamma_r_B12}
\end{equation}
where $d_1^2 = d_{\max}^2 - B \cos\Delta$.
Then the outage condition \(\gamma_r \leq \gamma_{\text{thr}}\) can be expressed as 
\begin{equation}
\frac{A \exp\left[ -\beta \left( \phi_u - \Delta \right) \right]}{d_1^2} \leq \gamma_{\text{thr}}.
\label{eq:exp_inequality}
\end{equation}
Taking the natural logarithm of both sides of \eqref{eq:exp_inequality} and then calculating, we have
\begin{equation}
\phi_u \geq \Delta + \frac{1}{\beta} \ln\left( \frac{A}{\gamma_{\text{thr}} d_1^2} \right)= \phi_{\text{th}}^{(B)}.
\end{equation}
The total outage angle is the part within the angular domain $[0, \theta]$ that is not less than $\phi_{\text{th}}^{(B)}$, expressed as 
\begin{equation}
\label{eq:L_constraint_C}
L^{(B)}(\rho) = \left[ \theta - \Delta - \frac{1}{\beta}\max\!\left(0,\; \ln\frac{A}{\gamma_{\text{thr}} d_1^2}\right) \right]^+.
\end{equation}

\textbf 3) $\phi_p^* = 0$
\begin{itemize}
    \item $D \le 1$, $\phi_u \le \phi_0^*$, and $\Delta >\phi_u$ 
\end{itemize}
  
 Substituting $\phi_p = 0$ into \eqref{gamma_r_arc}, we have
\begin{equation}
\gamma_r^{\text{opt}}= \frac{A}{d_{\max}^2 - B \cos\phi_u} .
\label{eq:gamma_ratio}
\end{equation}
Then the outage condition \(\gamma_r \leq \gamma_{\text{thr}}\) can be expressed as 
\begin{equation}
d_{\max}^2 - B \cos\phi_u \geq \frac{A}{\gamma_{\text{thr}}}.
\end{equation}
After calculation, we have
\begin{equation}
\cos\phi_u \leq \frac{d_{\max}^2}{B} - \frac{A}{B \gamma_{\text{thr}}}. 
\label{eq:cos_inequality}
\end{equation}
Considering that the PA is fixed at $\phi_p = 0$, and the user can establish LoS communication with the PA only when $\phi_u < \pi$, the range of $\phi_u$ is $[0, \pi)$. Thus \eqref{eq:cos_inequality} can be rewritten as
\begin{equation}
\phi_u \geq \arccos\left( \frac{d_{\max}^2}{B} - \frac{A}{B \gamma_{\text{thr}}} \right).
\label{eq:phi_threshold}
\end{equation}
Since the PA and the user must satisfy the LoS communication requirement, we have \(\phi_u \in [0, \min(\phi_0^*, \Delta)]\). Combined with \eqref{eq:phi_threshold}, the information outage range can be expressed as
\begin{equation}
L^{(C1)}(\rho) = \left[ \min(\phi_0^*, \Delta) - \arccos\left( \frac{d_{\max}^2}{B} - \frac{A}{B \gamma_{\text{thr}}} \right) \right]^+ .
\label{eq:L_function}
\end{equation}

\begin{itemize}
  \item $D > 1$, and $\Delta > \phi_{\mathrm{u}} $
    \end{itemize}
    
Referring to the derivation method of \eqref{eq:L_function}, the angular-domain outage measure can be expressed as
\begin{equation}
\label{eq:L_B1_B14}
L^{(C2)}(\rho) = \left[ \min(\theta, \Delta) - \arccos\left( \frac{d_{\max}^2}{B} - \frac{A}{B \gamma_{\text{thr}}} \right) \right]^+.
\end{equation}
Finally, we substitute \eqref{eq:L_function}, \eqref{eq:L_B1_B14} and \eqref{eq:L_constraint_C} into \eqref{eq:outage_radial}. 

The proof is completed.

\section{ PROOF OF PROPOSITION 4}
\renewcommand{\theequation}{C.\arabic{equation}}
\setcounter{equation}{0} %

Referring to the piecewise expression of the optimal PA position in \eqref{eq:optimal_phi_p_star}, the following three cases need to be considered when calculating \(J_(\rho)\)  in \eqref{J_angle_def}.

\textbf 1) Optimal PA Position without Blockage 

Substituting \eqref{eq:B1} into \eqref{J_angle_def},  \(J_(\rho)\) can be expressed as

\begin{equation}
\begin{split}
J^{(A)}(\rho) &= \int_{\phi_0^*}^{\theta} \log_2 \left( 1 + \frac{A e^{-\beta(\phi_u - \phi_0^*)}}{d_{\max}^2 - B \cos \phi_0^*} \right) d\phi_u. 
\end{split}
\label{eq:J_single_A_C1}
\end{equation}

Let $t = \phi_u - \phi_0^*$, then \eqref{eq:J_single_A_C1} can be rewritten as
\begin{equation}
\begin{split}
J^{(A)}(\rho) &= \int_{0}^{\theta - \phi_0^*} \log_2 \left( 1 + \frac{A e^{-\beta t}}{d_{\max}^2 - B \cos (\phi_0^*)} \right) dt.
\end{split}
\label{eq:JA}
\end{equation}
Convert \eqref{eq:JA} to natural logarithm, we have
\begin{equation}
\begin{split}
J^{(A)}(\rho) &= \frac{1}{\ln 2} \int_{0}^{\theta - \phi_0^*} \ln \left( 1 + \frac{A}{d_0^2} e^{-\beta t} \right) dt .
\end{split}
\end{equation}
Let $u = e^{-\beta t}$, we have
\begin{equation}
\begin{split}
J(\rho)^{(A)} &= \frac{1}{\beta \ln 2} \int_{e^{-\beta(\theta - \phi_0^*)}}^{1} \frac{\ln \left( 1 + \frac{A u}{d_0^2} \right)}{u} du. 
\end{split}
\label{eq:J_single_A_C4}
\end{equation}
Referring to the definition $\text{Li}_2(x) = -\int_{0}^{x} \frac{\ln(1-u)}{u} du$ of the dilogarithm function \cite[Eq. (6.254.1)]{gradshteyn2007table}, \eqref{eq:J_single_A_C4} can be transformed into
\begin{multline}
J^{(A)}(\rho) = \frac{1}{\beta \ln 2} \biggl[ -\text{Li}_2 \biggl( -\frac{A}{d_0^2} u \biggr) \biggr]_{e^{-\beta(\theta - \phi_0^*)}}^{1} \\
= \frac{1}{\beta \ln 2} \biggl[ \text{Li}_2 \biggl( -\frac{A}{d_0^2} e^{-\beta(\theta - \phi_0^*)} \biggr) - \text{Li}_2 \biggl( -\frac{A}{d_0^2} \biggr) \biggr]
\label{eq:Li2_result_C5}.
\end{multline}

\textbf 2) Optimal PA position with Blockage

Substituting \eqref{eq:gamma_r_B12} into \eqref{J_angle_def},  \(J_(\rho)\) can be expressed as
 \begin{equation}
\begin{aligned}
J^{(B)}(\rho) = \int_{\Delta}^{\theta} \log_2 \left( 1 + \frac{A e^{-\beta(\phi_u - \Delta)}}{d_1^2} \right) d\phi_u.
\end{aligned}
\end{equation}
Referring to the derivation process of  \eqref{eq:Li2_result_C5}, we have
\begin{equation}
\label{eq:Li2_result_C13}
\begin{aligned}
J^{(B)}(\rho) &= \frac{1}{\beta \ln 2} \biggl[ \operatorname{Li}_2 \biggl( -\frac{A}{d_1^2} e^{-\beta(\theta - \Delta)} \biggr) - \operatorname{Li}_2 \biggl( -\frac{A}{d_1^2} \biggr) \biggr].
\end{aligned}
\end{equation}

\textbf 3) $\phi_p^* = 0$

\begin{itemize}
  \item $D \le 1$, $\phi_u \le \phi_0^*$, and $\Delta > \phi_u$
    \end{itemize}

Substituting \eqref{eq:gamma_ratio}  into \eqref{J_angle_def}, \(J_(\rho)\) can be expressed as
\begin{equation}
\begin{split}
J^{(C)}(\rho) &= \int_{0}^{\min(\phi_0^*, \Delta)} \log_2 \left( 1 + \frac{A}{d_{\max}^2 - B \cos \phi_u} \right) d\phi_u .
\end{split}
\label{eq:JC}
\end{equation}
Convert \eqref{eq:JC} to natural logarithm, we have
\begin{equation}
\begin{split}
J^{(C)}(\rho) &= \frac{1}{\ln 2} \biggl[ \int_{0}^{\min(\phi_0^*, \Delta)} \ln(A + d_{\max}^2 - B \cos t) dt \\
&\qquad - \int_{0}^{\min(\phi_0^*, \Delta)} \ln(d_{\max}^2 - B \cos t) dt \biggr].
\end{split}
\label{eq:C9}
\end{equation}
Using the identity from \cite[Eq. (1.541)]{gradshteyn2007table} for the logarithmic integral with a cosine kernel, we obtain
\begin{equation}
\int_{0}^{\phi} \ln(1 - 2r \cos t + r^2) \, dt = -2 \operatorname{Im} \bigl[ \operatorname{Li}_2 (r e^{i\phi}) \bigr].
\end{equation}
On this basis, by transforming the first term of \eqref{eq:C9}, we have 
\begin{equation}
\begin{aligned}[b]
&\int_{0}^{\min(\phi_0^*, \Delta)} \ln(A + d_{\max}^2 - B \cos t) dt \\
&= \min(\phi_0^*, \Delta) \ln \frac{A + d_{\max}^2 + \sqrt{(A + d_{\max}^2)^2 - B^2}}{2} \\
&- 2 \operatorname{Im} \left[ \operatorname{Li}_2 \left( \frac{B e^{i\min(\phi_0^*, \Delta)}}{A + d_{\max}^2 + \sqrt{(A + d_{\max}^2)^2 - B^2}} \right) \right].
\end{aligned}
\end{equation}
Thus $J^{(C)}(\rho)$ is given by
\begin{equation}
\label{eq:Li2_result_C10}
\begin{aligned}
& J(\rho)^{(C1)} = J^{(C)}(\rho)\big|_{\phi = \min(\phi_0^*, \Delta)} - J^{(C)}(\rho)\big|_{\phi = 0} \\
&= \frac{1}{\ln 2} \Bigg[ \min(\phi_0^*, \Delta) \cdot \ln \frac{A + d_{\max}^2 + \sqrt{(A + d_{\max}^2)^2 - B^2}}{d_{\max}^2 + \sqrt{d_{\max}^4 - B^2}} \\
&\phantom{=} - 2\operatorname{Im} \operatorname{Li}_2\!\left( \frac{B}{A + d_{\max}^2 + \sqrt{(A + d_{\max}^2)^2 - B^2}} e^{i\min(\phi_0^*, \Delta)} \right) \\
&\phantom{=} + 2\operatorname{Im} \operatorname{Li}_2\!\left( \frac{B}{d_{\max}^2 + \sqrt{d_{\max}^4 - B^2}} e^{i\min(\phi_0^*, \Delta)} \right) \Bigg].
\end{aligned}
\end{equation}

\begin{itemize}
  \item $D > 1$, and $\Delta>  \phi_{\mathrm{u}} $
    \end{itemize}
    
Referring to the derivation method of \eqref{eq:Li2_result_C10}, $J^{(C)}(\rho)$ is given by
\begin{equation}
\label{eq:Li2_result_C11}
\begin{aligned}
& J(\rho)^{(C2)} = J^{(C)}(\rho)\big|_{\phi = \min(\theta, \Delta)} - J^{(C)}(\rho)\big|_{\phi = 0} \\
&= \frac{1}{\ln 2} \Bigg[ \min(\theta, \Delta) \cdot \ln \frac{A + d_{\max}^2 + \sqrt{(A + d_{\max}^2)^2 - B^2}}{d_{\max}^2 + \sqrt{d_{\max}^4 - B^2}} \\
&\phantom{=} - 2\operatorname{Im} \operatorname{Li}_2\!\left( \frac{B}{A + d_{\max}^2 + \sqrt{(A + d_{\max}^2)^2 - B^2}} e^{i\min(\theta, \Delta)} \right) \\
&\phantom{=} + 2\operatorname{Im} \operatorname{Li}_2\!\left( \frac{B}{d_{\max}^2 + \sqrt{d_{\max}^4 - B^2}} e^{i\min(\theta, \Delta)} \right) \Bigg].
\end{aligned}
\end{equation}
Finally, we substitute \eqref{eq:Li2_result_C5}, \eqref{eq:Li2_result_C10}, \eqref{eq:Li2_result_C11} and \eqref{eq:Li2_result_C13} into \eqref{J_angle_def}. 

The proof is completed.


\renewcommand\thesubsectiondis{\Roman{subsection}.}
\bibliographystyle{IEEEtran}
\bibliography{citation}

@article{Cai2025NextGenTRX,
  author={C. You and Y. Cai and Y. Liu and M. Di Renzo and T. M. Duman and A. Yener and A. L. Swindlehurst},
  title={{Next generation advanced transceiver technologies for 6G and beyond}},
  journal={IEEE J. Sel. Areas Commun.},
  volume={43},
  number={3},
  pages={582--627},
  month={Mar.},
  year={2025},
  doi={10.1109/JSAC.2025.3536557}
}

@article{cui2023near,
  title={{Near-field MIMO communications for 6G: Fundamentals, challenges, potentials, and future directions}},
  author={Cui, Mingyao and Wu, Zidong and Lu, Yu and Wei, Xianda and Dai, Linglong},
  journal={IEEE Commun. Mag.},
  volume={61},
  number={1},
  pages={40--46},
  year={2023},
  month={Jan.}
}

@article{Wang2023Road6G,
  author={C.-X. Wang and X. You and X. Gao and X. Zhu and Z. Li and C. Zhang and H. Wang and Y. Huang and Y. Chen and H. Haas and J. S. Thompson and E. G. Larsson and M. D. Renzo and W. Tong and P. Zhu and X. Shen and H. V. Poor and L. Hanzo},
  title={{On the road to 6G: Visions, requirements, key technologies, and testbeds}},
  journal={IEEE Commun. Surv. Tutorials},
  volume={25},
  number={2},
  pages={905--974},
  year={Feb. 2023}
}

@article{lu2024tutorial,
  title={{A tutorial on near-field XL-MIMO communications toward 6G}},
  author={Lu, Haiquan and Zeng, Yong and You, Changsheng and Han, Yu and Zhang, Jiayuan and Wang, Zhaoyang and Zhang, Rui},
  journal={IEEE Commun. Surv. Tutorials},
  volume={26},
  number={4},
  pages={2213--2257},
  year={Apr. 2024}
}

@article{cao2026reliable,
  title={{Reliable and secure wireless-powered communications via hybrid active-passive double-RIS}},
  author={Cao, K. and Wang, T. and Diamantoulakis, P. D. and Li, X. and Yuen, C. and Karagiannidis, G. K.},
  journal={IEEE J. Sel. Areas Commun.},
  volume={44},
  pages={3828--3844},
  month={Mar.},
  year={2026}
}

@article{cao2026self,
  title={{Self-sustainable active metasurface (SAM): Reliable and secure communications}},
  author={Cao, K. and Xu, S. and Diamantoulakis, P. D. and Yang, C. and Zheng, B. and Li, X. and Yuen, C.},
  journal={IEEE Trans. Wireless Commun.},
  volume={25},
  pages={12325--12340},
  year={Feb. 2026}
}

@article{chen2025double,
  title     = {{Double-RIS enabled physical layer security for wireless-powered communication systems over Rayleigh fading channels}},
  author    = {Chen, J. and Cao, K. and Ding, H. and Lv, L. and Ye, Y. and Chi, H. and Wang, T. and Yang, L.},
  journal   = {IEEE Trans. Commun.},
  year      = {2025},
  volume    = {73},
  number    = {10},
  pages     = {9517--9535},
  issn      = {0090-6778},
  doi       = {10.1109/TCOMM.2025.3548690}
}

@article{chen2025secure,
  title={{Secure wireless-powered zeRIS communications}},
  author={J. Chen and K. Cao and P. D. Diamantoulakis and L. Lv and L. Yang and H. Chi},
  journal={IEEE Trans. Wireless Commun.},
  volume={25},
  number={},
  pages={225--242},
  month={Jul.},
  year={2025},
  doi={10.1109/TWC.2025.3624521}
}

@ARTICLE{wang2026pentahedral,
  author={Wang, C. and Xu, H.-X. and Zhu, R. and Ding, H. and Li, B. and Wang, X. and Yang, C. and Lu, Y. and et al.},
  journal={IEEE Trans. Antennas Propag.},
  title={{3D-printed pentahedral polarization-division transmissive metadevice with versatile wavefronts}},
  year={2026},
  month={May},
  volume={74},
  number={5},
  pages={4915--4920},
  doi={10.1109/TAP.2026.3661559}
}

@article{Hong2026FASsurvey,
  author  = {H. Hong and K.-K. Wong and C.-B. Chae and H. Xu and X. Guo and F. R. Ghadi and Y. Chen and Y. Xu and B. Liu and K.-F. Tong and Y. Zhang},
  title   = {A contemporary survey on fluid antenna systems: Fundamentals and networking perspectives},
  journal = {IEEE Trans. Netw. Sci. Eng.},
  year    = {Sep. 2026},
  volume  = {13},
  pages   = {2305--2328},
  issn    = {2327-4697},
  doi     = {10.1109/TNSE.2025.3613225}
}

@article{New2024FASoutage,
  author={W. K. New and K.-K. Wong and H. Xu and K.-F. Tong and C.-B. Chae},
  title={{Fluid antenna system: New insights on outage probability and diversity gain}},
  journal={IEEE Trans. Wireless Commun.},
  volume={23}, number={1}, pages={128--140}, month={Jan.}, year={2024},
  doi={10.1109/TWC.2023.3297757}
}

@article{pang2026secure,
  title={{Secure NOMA empowered by movable antenna in SAGIN}},
  author={Pang, Mingliang and Wang, Chaowei and Zhang, Zhi and Zhang, Ping and He, Fangying and Xu, Lexi and Jiang, Fan and Quek, Tony Q. S.},
  journal={IEEE Wireless Commun.},
  volume={},
  year={Nov. 2025},
  pages={143--153},
  doi={10.1109/MWC.2025.3623589}
}

@article{Zhu2025Movable,
  author={L. Zhu and W. Ma and Z. Xiao and R. Zhang},
  title={{Movable antenna enabled near-field communications: Channel modeling and performance optimization}},
  journal={IEEE Trans. Commun.},
  volume={73},
  number={9},
  pages={7240--7256},
  month={Sep.},
  year={ 2025}
}

@article{Fukuda2022Pinching,
  author={A. Fukuda and H. Yamamoto and H. Okazaki and Y. Suzuki and K. Kawai},
  title={{Pinching antenna - using a dielectric waveguide as an antenna}},
  journal={NTT DOCOMO Tech. J.},
  volume={23},
  number={3},
  pages={5--12},
  month={Jan.},
  year={2022}
}

@article{DT2026HOW,
   author={D. Tyrovolas and S. A. Tegos and Y. Xiao and P. D. Diamantoulakis and S. Ioannidis and C. K. Liaskos and G. K. Karagiannidis and S. D. Asimonis},
  title={{How many pinching antennas are enough?}},
  journal={IEEE Internet Things J.},
  volume={13},
  number={10},
  pages={21994--22006},
  year={May 2026}
}

@article{Wang2025Modeling,
  author={Z. Wang and C. Ouyang and X. Mu and Y. Liu and Z. Ding},
  title={{Modeling and beamforming optimization for pinching-antenna systems}},
  journal={IEEE Trans. Commun.},
  volume={73},
  number={12},
  pages={13904--13919},
  month={Dec.},
  year={2025}
}

@article{yang2025pinching,
  author={Z. Yang and N. Wang and Y. Sun and Z. Ding and R. Schober and G. K. Karagiannidis and V. W. Wong and O. A. Dobre},
  title={{Pinching antennas: Principles, applications and challenges}},
  journal={IEEE Wireless Commun.},
  volume={33},
  number={2},
  pages={175--184},
  month={Apr.},
  year={2026}
}

@article{Ouyang2025SWANs,
  author={C. Ouyang and H. Jiang and Z. Wang and Y. Liu and Z. Ding},
  title={{Uplink and downlink communications in segmented waveguide-enabled pinching-antenna systems (SWANs)}},
  journal={IEEE Trans. Commun.},
  volume={74},
  number={},
  pages={3688--3703},
  month={Jan.},
  year={2026},
  doi={10.1109/TCOMM.2026.3665582}
}

@article{Tyrovolas2026Performance,
  author={D. Tyrovolas and S. A. Tegos and P. D. Diamantoulakis and S. Ioannidis and C. K. Liaskos and G. K. Karagiannidis},
  title={{Performance analysis of pinching-antenna systems}},
  journal={IEEE Trans. Cogn. Commun. Netw.},
  volume={12},
  pages={590--601},
  month={Apr.},
  year={2025}
}

@article{Pakravan2025AI,
  author={S. Pakravan and M. Ahmadzadeh and M. Zeng and X. Li and F. Fang},
  title={{AI-empowered resource allocation for wirelessly powered pinching-antenna systems}},
  journal={IEEE Trans. Veh. Technol.},
  volume={},
  number={},
  pages={1--6},
  month={Apr.},
  year={2026},
  doi={10.1109/TVT.2026.3684393},
  issn={0018-9545}
}

@misc{Zhang2025Directional,
  author={R. Zhang and Y. Shao and Y. Liu},
  title={Directional pinching-antenna systems},
  howpublished={arXiv preprint arXiv:2511.19133},
  year={2025}
}

@article{Shan2025Multicast,
  author={S. Shan and C. Ouyang and Y. Li and Y. Liu},
  title={{Multigroup multicast design for pinching-antenna systems: Waveguide-division or waveguide-multiplexing?}},
  journal={IEEE Trans. Commun.},
  volume={74},
  number={},
  pages={6228--6242},
  month={Mar.},
  year={2026},
  doi={10.1109/TCOMM.2026.3675461}
}

@article{Mao2025ISAC,
  author={W. Mao and Y. Lu and Y. Xu and B. Ai and O. A. Dobre and D. Niyato},
  title={{Multi-waveguide pinching antennas for ISAC}},
  journal={IEEE Trans. Wireless Commun.},
  volume={25},
  number={},
  pages={5846--5858},
  month={Oct.},
  year={2025},
  doi={10.1109/TWC.2025.3621316}
}

@article{Zhang2026iot,
  author  = {H. Zhang and B. Zhang and Y. Zhao and K. Yang and G. Zhang},
  title   = {Performance analysis of pinching-antenna-enabled internet of things systems},
  journal = {IEEE Internet Things J.},
  year    = { Dec. 2026},
  volume  = {13},
  number  = {7},
  pages   = {14499--14514},
  issn    = {2327-4662},
}

@article{Liu2026pin,
  author  = {Y. Liu and Z. Wang and X. Mu and C. Ouyang and X. Xu and Z. Ding},
  title   = {Pinching-antenna systems: Architecture designs, opportunities, and outlook},
  journal = {IEEE Commun. Mag.},
  year    = {Jan. 2026},
  volume  = {64},
  number  = {1},
  pages   = {190--196},
  issn    = {0163-6804},
}

@article{Xu2026Attenuation,
  author={Y. Xu and Z. Ding and R. Schober and T.-H. Chang},
  title={{Pinching-antenna systems with in-waveguide attenuation: Performance analysis and algorithm design}},
  journal={IEEE Trans. Wireless Commun.},
  volume={25},
  pages={14563--14578},
  year={Mar. 2026}
}

@article{Xu2026LoS,
  author={Y. Xu and Z. Ding and O. A. Dobre and T.-H. Chang},
  title={{Pinching-antenna system design with LoS blockage: Does in-waveguide attenuation matter?}},
  journal={IEEE Trans. Wireless Commun.},
  volume={25},
  number={},
  pages={17322--17337},
  month={May},
  year={2026},
  doi={10.1109/TWC.2026.3656870}
}

@article{Zhao2026Secure,
author={R. Zhao and S. Hu and D. Mishra and D. W. K. Ng},
title={{Robust and secure blockage-aware pinching antenna-assisted wireless communication}},
journal={IEEE Trans. Mobile Comput.},
volume={},
number={},
pages={1--18},
month={May},
year={2026},
doi={10.1109/TMC.2026.3695952},
issn={1536-1233}
}

@article{Sun2026RobustMulti,
  author={M. Sun and X. Mu and C. Ouyang and G. Shang and Y. Liu},
  title={{Robust beamforming for pinching-antenna systems-based multi-user communications}},
  journal={IEEE Wireless Commun. Lett.},
  volume={15},
  pages={2859--2863},
  month={Apr.},
  year={2026},
  issn={2162-2337},
}

@article{Sun2026RobustSingle,
  author={M. Sun and C. Ouyang and S. Wu and Y. Liu},
  title={{Robust beamforming for pinching-antenna systems}},
  journal={IEEE Trans. Veh. Technol.},
  volume={75},
  number={7},
  pages={15073--15078},
  year={2026},
  month={Jul.}
}

@ARTICLE{kozyreff2016dispersion,
  author={Kozyreff, Gregory and Acharyya, Nirmalendu},
  journal={Opt. Express},
  title={{Dispersion relations and bending losses of cylindrical and spherical shells, slabs, and slot waveguides}},
  year={2016},
  month={Dec.},
  volume={24},
  number={25},
  pages={28204--28220},
  doi={10.1364/OE.24.028204}
}

@ARTICLE{deck1998,
  author={Deck, R. T. and Mirkov, Mirko and Bagley, B. G.},
  journal={J. Lightwave Technol.},
  title={{Determination of bending losses in rectangular waveguides}},
  year={1998},
  month={Sep.},
  volume={16},
  number={9},
  pages={1703--1714},
  doi={10.1109/50.712240}
}

@book{gradshteyn2007table,
  title={{Table of Integrals, Series, and Products}},
  author={Gradshteyn, I. S. and Ryzhik, I. M.},
  address={New York, NY, USA},
  publisher={Academic},
  year={2007}
}
\end{document}